\documentclass[11pt]{article}

\everydisplay{\mathsurround 0pt} 
\usepackage[affil-it]{authblk}
\usepackage{amsmath}
\usepackage{amssymb}
\usepackage{amsthm}
\usepackage{color}
\usepackage[a4paper]{geometry}
\usepackage{xspace}
\usepackage[colorlinks,citecolor=blue,urlcolor=blue]{hyperref}
\usepackage{mathtools}
\usepackage[round]{natbib}
\usepackage[inline]{enumitem}
\usepackage{mathrsfs}
\usepackage{comment}
\usepackage{liumathdef}
\usepackage{algorithm}
\usepackage{algorithmic}
\usepackage{cancel}

\usepackage[capitalize]{cleveref}
\newenvironment{talign*}
 {\let\displaystyle\textstyle\csname align*\endcsname}
 {\endalign}
\newenvironment{talign}
 {\let\displaystyle\textstyle\csname align\endcsname}
 {\endalign}
\crefname{talign}{}{}
\crefname{equation}{}{}
\crefname{figure}{Figure}{Figures}

\newcommand{\R}{\mathbb{R}}
\newcommand{\N}{\mathbb{N}}

\newcommand{\bolds}{}

\newcommand{\beas}{\begin{eqnarray*}}
\newcommand{\enas}{\end{eqnarray*}}
\newcommand{\bea}{\begin{eqnarray}}
\newcommand{\ena}{\end{eqnarray}}

\newcommand{\RR}{{\mathbb{R}} }
\newcommand{\EE}{{\mathbb{E}} }

\DeclareMathOperator*{\arginf}{arg\,inf}

\newcommand{\gset}{\mathcal{G}}
\newcommand{\hset}{\mathcal{H}}
\newcommand{\gsteinset}[2]{\gset_{||\cdot||_{#1},Q_n,{#2}}}
\newcommand{\ksteinset}[1]{\gset_{#1}} 
\newcommand{\knorm}[1]{\norm{#1}_k}
\newcommand{\operator}[1]{\mathcal{T}{#1}} 
\newcommand{\oparg}[2]{(\operator{#1})({#2})} 
\newcommand{\opsub}[1]{\mathcal{T}_{#1}} 
\newcommand{\opsubarg}[3]{(\opsub{#1}{#2})({#3})} 
\newcommand{\stein}[3]{\mathcal{S}({#1},{#2},{#3})} 
\newcommand{\opsetstein}[1]{\stein{#1}{\operator{}}{\gset}} 
\newcommand{\ssd}[1]{\mathcal{SS}({#1},\operator{},\gset)} 
\newcommand{\ksd}{\mathrm{KSD}_k} 

\newcommand{\generator}[1]{\mathcal{A}{#1}} 
\newcommand{\genarg}[2]{(\generator{#1})({#2})} 
\newcommand{\supp}[1]{\mathrm{supp}({#1})}
\renewcommand{\grad}{\nabla}
\newcommand{\qtext}[1]{\quad\text{#1}\quad} 
 
\newcommand{\twonorm}[1]{\norm{#1}_2} 
\newcommand{\Unif}{\textnormal{Unif}}
\newcommand{\iid}{\textrm{i.i.d.}\xspace}
\newcommand{\dist}{\sim}
\newcommand{\distiid}{\overset{\textrm{\tiny\iid}}{\dist}}

\newcommand{\eps}{\epsilon}

\def\reals{\mathbb{R}} 

\def\Ito{It\^o\xspace}

\def\Matern{Mat\'ern\xspace}

\usepackage[colorinlistoftodos]{todonotes}

\newtheorem{theorem}{Theorem}

\newtheorem{example}{Example}
\newtheorem{rem}{Remark}

\let\OLDthebibliography\thebibliography
\renewcommand\thebibliography[1]{
  \OLDthebibliography{#1}
  \setlength{\parskip}{3pt}
  \setlength{\itemsep}{0pt plus 0.3ex}
}

\title{Stein's Method Meets Computational Statistics:\\ A Review of Some Recent Developments}

\author{Andreas Anastasiou$^1$,
  Alessandro Barp$^2$, Fran\c cois-Xavier Briol$^3$, Bruno Ebner$^{4}$, Robert E. Gaunt$^5$, Fatemeh Ghaderinezhad$^6$, Jackson Gorham$^7$, Arthur Gretton$^{3}$, Christophe Ley$^{8}$\footnote{Correspondance should be addressed
    to \texttt{christophe.ley@ugent.be}.}, Qiang Liu$^{9}$, Lester Mackey$^{10}$, Chris. J. Oates$^{11}$, Gesine Reinert$^{12}$, Yvik Swan$^{13}$ \\

\small $^1$University of Cyprus,  Cyprus;
\small $^2$University of Cambridge, UK; 
\small $^3$University College London, UK; 
\small $^4$Karlsruhe Institute of Technology, Germany; 
\small $^5$The University of Manchester, UK;
\small $^6$Ghent University, Belgium;   
\small $^7$Whisper.ai, Inc., US;
\small $^8$University of Luxembourg, Luxembourg;
\small $^{9}$The University of Texas at Austin, US
\small $^{10}$Microsoft Research New England, US;
\small $^{11}$Newcastle University, UK;
\small $^{12}$University of Oxford, UK;
\small $^{13}$Universit\'e  Libre de Bruxelles, Belgium
}

\date{}

\begin{document}

\maketitle

\begin{abstract}

\indent Stein's method compares probability distributions through the study of a class of linear operators called Stein operators. 
While mainly studied in  probability and used to underpin theoretical statistics, Stein's method has led to significant advances in computational statistics in recent
years.  The goal of this survey is to bring together
some of these recent developments  and, in doing so,
to stimulate further research into the successful field of Stein's
method and statistics.  
The topics we discuss include 
tools to benchmark and compare sampling methods such as approximate Markov chain Monte Carlo, deterministic alternatives to sampling methods,  control variate techniques, parameter estimation and goodness-of-fit testing. 
\end{abstract}


{\it MSC2020 subject classifications}: Primary 62-02; Secondary 62-E99, 62-08

\section{Introduction}
\label{sec:intro}

\medskip Stein's method was 
introduced by Charles Stein in the early 1970s \citep{Stein72} for distributional comparisons to the normal distribution. At the foundation of Stein's method lies a characterizing equation for the normal distribution. This equation is also a cornerstone in Stein's unbiased estimator of risk \citep{stein1981estimation} and James-Stein shrinkage estimators \citep{stein1956inadmissibility, james1961estimation}; see  \citet{fathi2020relaxing} for a joined-up view. The latter paper also exploited these connections with Stein’s method to propose and analyse new estimators in a non-Gaussian setting. Here we concentrate on Stein's method for distributional comparisons.

Originally developed for normal approximation, the method was extended first to Poisson approximation by \cite{chen1975poisson}, then by a growing community to a growing collection of approximation problems including beta, binomial, gamma, Kummer-U, multinomial, variance-gamma, Wishart, and many more. Stein's method has proved powerful in particular for deriving explicit bounds on distributional distances even when the underlying random elements are structures with dependence. Moreover, it thrives 
when the target distribution is known only up to a normalizing constant. Comprehensive introductions to the theory and its applications are available in the monographs \citet{stein1986approximate,barbour1992poisson,chen2010normal, nourdin2012normal, arras2019stein}.  We also refer to the lecture notes of \cite{DH04} and the surveys  of \citet{ross,chatterjee2014short,barbourchen14,ley2017stein}. The websites \href{https://sites.google.com/site/malliavinstein}{https://sites.google.com/site/malliavinstein} and \href{https://sites.google.com/site/steinsmethod}{https://sites.google.com/site/steinsmethod} provide regularly updated lists of references.

Over the past few decades, Stein's method has had substantial interactions with other mathematical fields, such as Malliavin calculus, information theory, functional analysis, dynamical systems and stochastic geometry. Some examples of
applications of Stein’s method in theoretical statistics are as follows. \citet{stein2004use} employed the method  for the analysis of sample quality in simulations, \citet{HolRei04}
developed a 
bootstrap method for network data which is analysed via empirical processes, \cite{Shao05} obtained a Berry-Esseen bound for Student's $t$-statistic. 
Applications to  self-normalized limit theorems and 
false discovery rates in simultaneous tests are surveyed 
in \cite{shao2010stein}. In \cite{shao2016stein}, an overview on the use of randomised concentration inequalities in Stein's method for nonlinear statistics is provided. \cite{Lippert}; \cite{reinert2009alignment} 
utilized the method to 
prove that there were flaws in then commonly used statistics for alignment-free sequence comparison,  and subsequently introduce two new sequence comparison statistics which avoid these flaws. This list is by no means exhaustive, but has the goal to give the reader a first taste of the versatile usage of Stein's method in statistics.

Starting around 2015, these early and ongoing successes of Stein's method in theoretical statistics
have 
attracted the attention of researchers from computational statistics and machine learning.
 Indeed, the fact that target distributions only need to be known up to a normalizing constant for Stein's method to apply has sparked considerable interest in these domains.  Here, ingredients from Stein's method such as so-called \emph{Stein discrepancies} have been used to develop new methodological procedures based on Stein operators. The aim of this paper is to cover various (clearly not all) developments that took place in 
 computational statistics and machine learning 
 since around 2015; the choice of topics is biased by the research interests of the contributors. Related developments in applications of Stein's method in theoretical statistics are also included.  By this survey, we 
 wish to  bring Stein's method and its different ingredients to the attention of the broad statistical community in order to further foster this fertile research domain. 
 
This paper starts with a succinct introductory section on Stein's method (Section \ref{steinsection}), followed by Section \ref{sec:comp_stein_discrepancies} which discusses the practical issue of computing Stein discrepancies. Section~\ref{sec:B} presents various new  statistical and machine learning procedures for assessing sample quality as well as constructing sample approximations and improving Monte Carlo integration which are obtained by means of Stein method ingredients.
Section~\ref{sec:goodness-of-fit} details 
 new developments 
for testing goodness-of-fit which are based on Stein's method,  and discusses novel insights into existing inferential procedures such as the quality of asymptotic approximation of estimators and test statistics as well as the impact of the prior choice in Bayesian statistics. 
 Section \ref{sec:conclu} then provides some summarizing conclusions.

 \section{The Basic Ingredients of Stein's Method}\label{steinsection}

Stein's method provides a collection of tools permitting to  quantify the dissimilarity between  probability distributions. The method has many components, not all of which are pertinent to the present survey. The purpose of this introductory  section is to provide a succinct overview of the basic ingredients  which shall be of use in the rest of the paper. 

First we fix some notation. The distribution of a random quantity $X$ is  denoted by $\mathcal{L}(X)$. Expectations with respect to a probability distribution $Q$ are denoted by $\EE_{X \sim Q}$; sometimes the subscript is omitted when the context is clear.   The space $L^p(Q)$ denotes the set of functions such that $\EE_{X \sim Q}[|f^p(X)|]$ is finite.

The function $\mathbb{I}_A(x)$ is the indicator function of  $x \in A$, taking the value 1 if $x \in A$ and 0 otherwise. 
For $\R^d$-valued functions $f$ and $g$, the notation 
$\langle f, g \rangle$ denotes the inner product; if $f$ and $g$ are matrix-valued, it denotes the Hilbert-Schmidt inner product. The notation $C^k(\R^d)$ denotes functions in $\R^d$ that are $k$ times continuously differentiable.
The norm $| \cdot |$ is the absolute value, $|| \cdot ||_2$ the Euclidean norm and $|| \cdot ||_\infty$ denotes the supremum norm. The operator $\nabla$ denotes the gradient operator; the gradient of a smooth function {$v: \R^d \to \R$ is the vector valued function $\grad v$} with entries
$ (\grad v)_{{i}} = \partial_{i} v$, ${i}=1, \ldots, d$, by convention viewed as column vector. For a $d$-vector-valued function $\mathbf{v}: \R^d \rightarrow {\R^d}$
with components $v_j, j = 1, \ldots, d$, the divergence is $\mathrm{div} (v) = \nabla^\intercal v = \sum_{i=1}^d \partial_i v_i( x).$
For a vector or a matrix, the superscript $\intercal$ stands for the transpose; this also applies for vector- or matrix-valued operators. Finally, by convention, $0/0=0$.

\subsection{Stein Operators, Stein Discrepancies and Stein Equations}
\label{sec:stein-oper-discr}

The starting point of Stein's method for a target probability distribution $P$ on some set $\mathcal{X}$ consists in identifying a linear  operator $\operator{}$ acting on a set $\mathcal{G}(\operator{})$ of functions on $\mathcal{X}$
 such that, for any other probability measure $Q$ on $\mathcal{X}$, it holds that  
\begin{talign}
  \label{eq:Steinchar}
  Q = P \mbox{ if and only if  } \mathbb{E}_{X \sim Q} [
  \oparg{g}{X}] = 0 \mbox{ for all } g \in \mathcal{G}(\operator{}).
\end{talign}
Such an operator  $\operator{}$ is called a \emph{Stein operator}, the  collection $\mathcal{G}(\operator{})$ of functions for which $\mathbb{E}_{X \sim P} [\oparg{g}{X}] = 0 $  is called a \emph{Stein class}, and equivalence  \eqref{eq:Steinchar} is called a \emph{Stein characterization}. In many cases the characterizing nature of the operator is superfluous, and we only need to require that a \emph{Stein identity} for $P$ is satisfied, namely that $\mathbb{E}_{X \sim P} [\oparg{g}{X}] = 0 $ for all $g \in \mathcal{G}(\mathcal T)$. Through a Stein identity, we only have a guarantee that the expectations
taken against $P$ vanish, but they could also be zero when taken against some $Q \neq P$.

We will discuss the topic of choosing Stein operators in  Section~\ref{sec:choossteinop}. At this stage let us suppose that  we are given a {characterizing} Stein operator $\mathcal T$ with Stein class $\mathcal G(\mathcal T)$. Then, for any \emph{Stein set} $\mathcal G \subset \mathcal G(\mathcal T)$, one may define a dissimilarity measure  as
\begin{talign}
  \label{eq:SteinDisc}
\opsetstein{Q} =    \sup_{g \in
  \mathcal{G}}  \left\|\mathbb{E}_{X \sim Q} [
  \oparg{g}{X}]\right\|^*
\end{talign}
for some appropriate norm $\| \cdot\|^*$. By construction, if $\opsetstein{Q} \neq 0$, then $Q \neq P$ and, if $\mathcal{G}$ is sufficiently large, then $\opsetstein{Q} = 0$ also implies  $Q = P$.
\citet{GorhamMa15} call the quantity \cref{eq:SteinDisc} a \emph{Stein discrepancy}  
(in contrast to the use of the term in \citet{ledoux2015stein}).  
If the Stein
operator $\operator{}$ and the {Stein set}
$\mathcal{G} \subset \mathcal{G}(\operator{})$ are well-chosen, the Stein discrepancy $\opsetstein{Q}$ ought to capture some aspect of the dissimilarity between $P$ and $Q$. Part of the magic of Stein's method lies in the fact that there are numerous combinations of target distribution $P$ and approximating distribution $Q$ for which one can identify operators $\operator{}$ and sets $\mathcal{G}$ ensuring that the quantity $\mathcal{S} (Q, \operator{}, \mathcal{G}) $ is both tractable \emph{and} relevant.

As illustration, we now give an example of Stein discrepancy 
for quantifying the dissimilarity between any probability distribution $Q$ on $\R^d$ and the normal
distribution.

\begin{example}[Stein operator and discrepancy for the multivariate
  normal distribution]\label{ex:steingauss}
  Let $\Sigma$ be a $d \times d$ positive definite matrix; denote by  $\mathrm{N}_d(0,\Sigma)$ the centered multivariate normal with covariance $\Sigma$.  Let $g : \R^d \to \R$ {be almost differentiable}, i.e.\ possess a gradient $\nabla g : \mathbb{R}^d \to \mathbb{R}^d$ such that, for all $z \in \mathbb{R}^d$, $g(x+z) - g(x) = \int_0^1 \left\langle z, \nabla g(x+ tz)  \right\rangle \mathrm{d}t$ for almost all $x \in \mathbb{R}^d$.  Suppose furthermore that $\nabla g \in L^1(\mathrm{N}_d(0,\Sigma))$.
  Then 
   \begin{talign*}\mathbb{E}_{X \sim \mathrm{N}_d(0, \Sigma)} \left[\Sigma  \nabla g(X) - X g(X)\right] =0,
  \end{talign*}
  see for example  \citet{stein1981estimation} (for $\Sigma $ the identity matrix). We deduce that the first order    differential operator 
  \begin{talign} \label{normalop}
  \oparg{g}{x}  =  \Sigma  \nabla g(x) - x g(x)
  \end{talign} 
  is a Stein operator for $\mathrm{N}_d(0, \Sigma)$ acting on the Stein class $\mathcal{G}(\operator{})$ of all  almost differentiable functions with (almost everywhere) gradient $\nabla g \in L^1(N_d(0, \Sigma))$.  This leads to 
\begin{talign}\label{normaldis} 
\opsetstein{Q} = \sup_{g \in
    \mathcal{G}} \left\| \mathbb{E}_{X \sim Q} \left[   
      \Sigma  \nabla g(X) -
      X g(X) \right]\right\|_2,
\end{talign}
for any $\mathcal{G} \subset \mathcal{G}(\operator{})$.
\end{example}

Of course it remains to ensure that the dissimilarity measures herewith obtained actually capture relevant aspects of the dissimilarity between $P$ and $Q$. Classically, there are many ways to determine discrepancies between probability measures, see for example \cite{gibbs2002choosing,rachev2013methods}. In this survey, and in much of the literature on Stein's method, the focus is on  distances known as \emph{integral probability metrics} (IPMs, for short) \citep{Muller97,  zolotarev1984probability}, which are defined as 
\begin{talign}\label{eq:IPM}
d_{\mathcal{H}}(P, Q) :=    \sup_{h \in \mathcal{H}}
   \left| \mathbb{E}_{X \sim P}[h(X)]-  \mathbb{E}_{X \sim Q} [h(X)] \right|
\end{talign}
for some class of real-valued measurable test functions
$\mathcal{H} \subset L^1(P) \cap L^1(Q)$. When $d_\mathcal{H}$ is a distance on the set of probability measures on $\mathcal X$ then $\mathcal H$ is called \emph{measure determining}. 

\begin{rem} \label{idmlist} Different choices
of $\mathcal{H}$ give rise to different {IPMs},  including: 
  \begin{enumerate} 
  \item   the \emph{Kolmogorov distance}: $d_{\mathrm{Kol}}(P, Q)$, which is the IPM induced by the set of test functions $\mathcal{H}_{\mathrm{Kol}} = \left\{ \mathbb{I}_{(-\infty, x]}(\cdot) \, : \, x \in \R^d \right\}$ (indicators of bottom left quadrants); 
\item the \emph{$L^1$-Wasserstein distance} (also known as the Kantorovich-Rubinstein or earth-mover's distance): $d_{\mathrm{W}}(P, Q)$, which  is the IPM induced by the set of test functions $\mathcal{H}_{\mathrm{W}} = \{h:\mathbb{R}^d\to\mathbb{R}\,:\, \sup_{x\neq y\in\mathbb{R}^d} |h(x)-h(y)|/\|x-y\|_2 \le 1\}$ (functions with Lipschitz constant at most 1); 
\item the \emph{bounded Wasserstein distance} (also known as the Dudley or bounded Lipschitz metric): $ d_{\mathrm{bW}}(P, Q)$, which is the IPM induced by the set of test functions $\hset_\mathrm{bW}$ which collects the bounded functions in $\mathcal{H}_{\mathrm{W}}$;
\item the \emph{maximum mean discrepancy}: $ d_{k}(P, Q)$, which is the IPM induced by the set of test functions $\hset_k$, the unit-ball of some reproducing kernel Hilbert space \citep{Berlinet2004} associated with kernel $k$. This case will be discussed extensively in \Cref{sec:comp_stein_discrepancies}.
\end{enumerate}
\end{rem} 

To see the connection between IPMs $d_{\mathcal{H}}$ and Stein discrepancies $\mathcal{S}$, an additional ingredient enters the picture: the \emph{Stein equation}. Given $P$ the target distribution with Stein operator $\operator{}$ and Stein class
$\mathcal{G}(\operator{})$, and given $\mathcal{H} \subset L^1(P)$ a measure-determining  class of test functions, the \emph{Stein equation} for $h \in \mathcal{H}$ is the functional equation 
\begin{talign}\label{eq:steinequ}
  \oparg{g}{x} = h(x) - \mathbb{E}_{X \sim P}[h(X)]
\end{talign}
evaluated over $x \in \mathcal X$, with solution $g = g(h) := \mathcal{L}h\in \mathcal{G}$, if it exists. Assuming that this solution exists {for all $h \in \mathcal{H}$}, it follows that $\mathbb{E}_{X \sim Q}[h(X)] - \mathbb{E}_{X \sim P}[h(X)] = \mathbb{E}_{X \sim Q} [ \oparg{(\mathcal{L}h)}{X}]$ so 
\begin{talign*}
 d_{\mathcal{H}}(P, Q)  =    \sup_{h\in\mathcal{H}}
 |\mathbb{E}_{X \sim P}[h(X)] - \mathbb{E}_{X \sim Q}[h(X)]|   = \mathcal{S}(Q, \operator{},
 \mathcal{L}\mathcal{H})
\end{talign*}
with $\mathcal{L}\mathcal{H}$ the Stein set collecting  all solutions $\mathcal L h$ to the Stein equation 
\eqref{eq:steinequ} with $h \in \mathcal H$.   Existence of a solution to the Stein equation depends on the properties of the target measure $P$, of the Stein operator $\mathcal T$, and of the Stein class  $\mathcal G(\mathcal T)$. In many cases, existence of these solutions is guaranteed and {the}  IPMs listed in Remark \ref{idmlist}  can be rewritten as Stein discrepancies whose underlying Stein set $\mathcal{L}\mathcal{H}$ depends on the measure $P$ characterized by $\operator{}$ through \eqref{eq:steinequ}.

Often, bounding $\mathbb{E}_{X \sim Q}[h(X)] - \mathbb{E}_{X \sim P}[h(X)]$ through bounding  $ \mathbb{E}_{X \sim Q} [ \oparg{(\mathcal{L}h)}{X}]$ is advantageous as the latter only requires integrating under $Q$; the properties of $P$ have been encoded in the Stein operator and Stein class. Commonly used approaches for bounding Stein discrepancies are coupling techniques \citep{barbour1992poisson, reinert98, chen2010normal,  ross}, the Malliavin-Stein method \citep{nourdin2012normal} and comparison of Stein operators  \citep{Holmes2004,ley2017stein,mijoule2021stein}; here the references only serve as pointers and the list is certainly not complete. In the context of theoretical statistics,  IPM-based Stein discrepancies have been used  for investigating finite-sample performance of statistical estimators  with intractable exact distribution and known asymptotic behavior (here, thus, $Q$ is the exact distribution of some statistical procedure, and $P$ its asymptotic distribution).   An overview of some of these applications   is provided in {Section \ref{sec:goodness-of-fit}}.

In order to bound  $ \mathbb{E}_{X \sim Q} [ \oparg{(\mathcal{L}h)}{X}]$, suitable bounds on the solutions $\mathcal Lh$ of the Stein equation, as well as certain lower order derivatives {or differences} of the solution, are usually required (although sometimes weak solutions of an appropriate equation suffice, see  \citet{courtade2019existence, mijoule2021stein}). Bounds on the solution are often referred to as \emph{Stein factors}. Determining Stein factors has attracted attention in recent years. Of the many available references, we single out \cite{mackey2016multivariate, fang2019multivariate} where bounds are obtained for operators given in the setting of Example \ref{ex:steinopdiffusions} under assumptions of log-concavity, and  
\cite{GorhamDuVoMa19}, where Stein factors are obtained under the weaker assumption of \emph{integrable Wasserstein decay}. An overview for continuous distributions is given in \cite{mijoule2021stein}.

In this section, we have kept $\H$, or equivalently $d_\H$, {mainly} general, so that the task of deriving a Stein equation and bounds on Stein factors can be presented in a form which applies to any of the IPMs {in Remark \ref{idmlist}.}

\subsection{Choosing Stein Operators}
\label{sec:choossteinop}

 When tackling Stein's method for a general target via a Stein discrepancy $\mathcal{S}(Q,
  \operator{}, \mathcal G)$, it is important to {first} choose $\mathcal T$ and $\mathcal G$ in a way which ensures relevance and  tractability of the resulting metric or Stein discrepancy. For many target distributions, 
 such useful Stein operators and Stein sets are readily available from the literature.  One of the advantages of Stein's method, however, is that for a given $P$ there is in principle full  freedom of choice in the operator $\mathcal T$ and Stein set $\mathcal G$, and in particular  no need to restrict  to the operators from the literature nor Stein sets obtained from Stein equations. 
   
 Here we shall {mainly} concentrate on two approaches {for choosing a Stein operator}, called the generator approach (which dates back to \citet{Barbour88, Barbour90} {and \citet{Gotze91}}) and the density approach (which dates back to \citet{stein2004use}). These are not the only available approaches (see for example \citet{reinert2005three}) and we conclude the section with a brief pointer to other techniques.

\subsubsection{Stein Operators via the Generator Approach} \label{sec:generator_approach}

We first describe the \emph{generator approach}, which we  present for {a given target $P$ on} $\X = \R^d$. Given a Markov process with sufficient regularity $(Z_t)_{t\ge 0}$ (namely, a Feller process \citep[Lemma 8.1.4]{Oksendal2013}) with invariant measure $P$, the \emph{infinitesimal generator} $\generator{}$ of the process  given by
\begin{talign*}
\genarg{u}{x} = \lim_{t\to 0} \frac{1}{t} ({\mathbb{E}[u(Z_t) \mid Z_0 = x] - u(x)})
\end{talign*}
satisfies the property that $\mathbb{E}_{Z \sim P}[\genarg{u}{Z}] = 0$ for all $u:\mathbb{R}^d\to\mathbb{R}$ in the domain of $\generator{}$.
\cite{Barbour88,Barbour90} and \cite{Gotze91} exploited this fact to provide both a Stein operator and a Stein class for all targets $P$ that are invariant measures of sufficiently regular Markov processes, to analyse multivariate distributions via Stein's method. 

\citet{GorhamDuVoMa19} detailed the generator approach for a wide range of distributions of interest by using operators induced by \emph{\Ito diffusions}. An \Ito diffusion \cite[Def.~7.1.1]{Oksendal2013} with starting point $x\in\mathbb{R}^d$, Lipschitz \emph{drift coefficient} $b : \mathbb{R}^d \to \mathbb{R}^d$, and Lipschitz \emph{diffusion coefficient} $\sigma : \mathbb{R}^d \to \mathbb{R}^{d\times m}$ is a stochastic process $(Z_{t,x})_{t\geq0}$ solving the \Ito stochastic differential equation 
\begin{talign}\label{eqn:diffusion}
\mathrm{d} Z_{t,x} = b(Z_{t,x})\, \mathrm{d}t + \sigma(Z_{t,x})\,\mathrm{d}W_t\text{ with } Z_{0,x}=x\in\mathbb{R}^d,
\end{talign}
where $(W_t)_{t\geq 0}$ is a $m$-dimensional Brownian motion. It is known (see, e.g.,\ \citet[Thm.~2]{GorhamDuVoMa19} and \citet[Thm.~19]{barp2020bracket}) that equation  \eqref{eqn:diffusion} will have invariant measure
$P$ with density $p$ {which is positive and differentiable} if and only if $b(x) = \langle\nabla, p(x)[\sigma(x)\sigma(x)^{\intercal} + c(x)]\rangle/2p(x)$ where  the \emph{stream coefficient}
$c:\mathbb{R}^d\to\mathbb{R}^{d\times d}$ is some differentiable skew-symmetric valued function.
\citet{GorhamDuVoMa19} proposed the first order \emph{diffusion Stein operator}
\begin{talign}\label{eqn:diffusion-stein-operator}
(\operator{}g)(x) = \frac{1}{p(x)}\langle \nabla, p(x)[(\sigma(x)\sigma(x)^{\intercal} + c(x)]g(x)\rangle,
\end{talign}
based on the diffusion's second order infinitesimal generator $\generator{u} = \operator{}(\nabla u/2)$. Under regularity conditions, the definition in equation \eqref{eqn:diffusion-stein-operator} yields an infinite collection of Stein operators for a given target $P$, parametrized by the choice of $\sigma$ and $c$.

\begin{example}[The  Langevin Stein operator on $\mathbb{R}^d$] \label{ex:steinopdiffusions}
As a concrete example, \citet{GorhamMa15,mackey2016multivariate} consider the case where $\sigma \equiv I_d$ and $c \equiv 0$, which corresponds to the overdamped Langevin diffusion. Assuming $\mathbb{E}_{X \sim P}[\|\nabla \log p(X) \|_2] < \infty$, this induces the Langevin Stein operator
\begin{talign}
(\operator{}g)(x) = \langle \nabla \log p(x), g(x)\rangle + \langle
\nabla, g(x)\rangle. \label{eq: Langevin Stein operator}
\end{talign}
The corresponding Stein discrepancies from equation \eqref{eq:SteinDisc} are often called {\emph{Langevin Stein discrepancies}}.
\end{example}

\subsubsection{Stein Operators via the Density Approach}
\label{sec:steinopdensap}

The \emph{density approach} was pioneered in \cite{stein2004use} for univariate distributions, and has since then been generalized in multiple directions, see for example \cite{yang2018goodness, mijoule2021stein}. Given a probability measure  $P$ on a set $\mathcal X$ with  density function  (with respect to some dominating measure) $p : \mathcal X \to \mathbb{R}^+$,  consider   operators of  the form $g \mapsto  {\mathcal{D}(g(x)p(x))}/{p(x)}$, where  $\mathcal D$ is a linear operator with domain $\mathrm{dom}(\mathcal D)$. Collecting into the  class $\mathcal G$ all functions $g$ on $\mathcal X$ such that $x \mapsto p(x) g(x) \in \mathrm{dom}(\mathcal D)$ and $\int_{\mathcal{X}} \mathcal{D}(g(x)p(x)) \,\mathrm{d}x = 0$,  the \emph{$\mathcal{D}$- density}, or for short, \emph{density} Stein operator of the density approach for $p$ is 
\begin{talign*}
g \mapsto (\operator{}g)(x) = \frac{\mathcal{D}(g(x)p(x))}{p(x)}
\end{talign*}
with Stein class $\mathcal G(\mathcal T) = \mathcal G$. By construction, this operator satisfies $\mathbb{E}_{X \sim P} [ (\operator{}g)(X)]=0$ for all $g \in \mathcal{G}(\mathcal{T})$. The following example illustrates the approach for univariate distributions with interval support. 

\begin{example}[Density {operators} for the exponential distribution] Fix $d=1$ and {consider as target} $P$ the exponential distribution with density  function  $p(x) = \lambda \mathrm{e}^{- \lambda x}\mathbb{I}_{[0, \infty)}(x)$, for $\lambda>0$. A natural choice of $\mathcal D$ is  $\mathcal D f(x) = f'(x)$ the usual almost everywhere derivative. If $(g p)'$ is integrable on $\R^+$, then  $\int_0^\infty (g(x) p(x))' \,\mathrm{d}x = \lim_{x \to \infty} g(x) p(x) - \lambda g(0)$. 
The corresponding density operator is therefore 
\begin{talign*}
(\mathcal{T} g)(x) = \frac{(g(x)p(x))'}{p(x)} = g'(x) - \lambda g(x), \qquad x \in \R^+
\end{talign*}
acting on the Stein class of functions $g$ such that $(g p)'$ is integrable on $\R^+$ and  $\lim_{x \to \infty} g(x) p(x) = \lambda g(0)$. Clearly all functions $g(x) = x g_0(x)$  such that  $\lim_{x \to \infty} x g_0(x) \mathrm{e}^{- \lambda x} =0$ belong to $\mathcal{G}(\mathcal T)$. Denoting $\tilde{\mathcal G}$ the collection of functions of this form, we reap a second operator for the exponential given by
\begin{talign*}
(\mathcal T_1 g_0)(x) = \frac{(x g_0(x) \mathrm{e}^{-\lambda x})'}{\mathrm{e}^{-\lambda x}}  = x g_0'(x) + (1-\lambda x) g_0(x)
\end{talign*}
acting on the (restricted) Stein class $\tilde{\mathcal G}$. The advantage of the latter operator over the former is that it does not require any implicit boundary assumptions on the test functions.
  
  Since the exponential density is also a parametric scale family in its parameter $\lambda>0$, another natural derivative in this context is  $\mathcal D f(x; \lambda) = \frac{\mathrm{d}}{\mathrm{d} \lambda} f(x; \lambda)$ for all functions $f(x; \lambda)$ of the form $f(x; \lambda) = \lambda f_0( \lambda x)$ for some $f_0$. This leads to 
  \begin{talign*}
  (\mathcal T_2 g)(x) = \frac{\frac{\mathrm{d}}{\mathrm{d} \lambda} ( \lambda 
  g(\lambda x) \mathrm{e}^{-\lambda x})}{ (\lambda \mathrm{e}^{-\lambda x})} =  x g'( \lambda x)+(\frac{1}{\lambda} -x)g(\lambda x),
  \end{talign*}
 with no boundary assumptions on $g$ since $\mathbb E_{X \sim \mathrm{Exp}(\lambda)} [(\mathcal T_2 g)(X)] = \frac{\mathrm{d}}{\mathrm{d} \lambda} \left(\int_0^\infty g(u) \mathrm{e}^{-  u}\, \mathrm{d}u\right) = 0$ for all $g \in L^1(\mathrm{Exp}(1))$. 
\end{example}

Many choices of operator $\mathcal D$ lead to Stein operators. Moreover, using appropriate product rules, Stein operators can be tailored for the specifics of the problem at hand. This process is called \emph{standardizing the Stein operator}, see \cite{kattumannil2009stein} and \cite{goldstein2013stein}.

The density approach and the generator approach are by no means the only methods for obtaining Stein operators. Other popular approaches include couplings (\citet{chen2010stein}), orthogonal polynomials (\citet{goldstein2005distributional}), a  perturbation approach (\citet{barbour1999poisson}),  an ODE approach (\citet{gauntpn17}), and characteristic functions (\citet{tikhomirov1980rate}; \citet{arras2019stein}). 

\subsubsection{Some General Remarks on Stein Operators}
\label{sn:genremar}

A Stein operator can often be found even when the density of the target distribution is not available in closed form, which will be particularly useful for applications in statistics. In this context we highlight two classes of important problems:

\paragraph{Bayesian Computation} In Bayesian statistics, usually the posterior distribution is known only in an unnormalized form. This is not a hindrance for Stein's method, see  \cite{LRS17}. Take for example the Langevin Stein operator of Example \ref{ex:steinopdiffusions}: $(\operator{}g)(x) = \langle \nabla \log p(x), g(x)\rangle + \langle
\nabla, g(x)\rangle$. Any function of the form $(\operator{}g)$ can be evaluated pointwise provided that $\nabla \log p$ can be evaluated, which is often a reasonable requirement. In particular, this does not require knowledge of the normalizing constant of $p$, since if $p=\tilde{p}/C$ for $C >0$, then 
$\nabla \log p = \nabla \log \tilde{p} -\nabla \log C = \nabla \log \tilde{p}$. In fact $\nabla \log p$ is usually the basis of gradient-based Markov chain Monte Carlo algorithms to sample from posterior distributions.
Illustrations of this principle can be found in   \citet{GorhamDuVoMa19} and \citet{mijoule2021stein}, for instance.

\paragraph{Intractable Likelihood}
A second example includes models in which the likelihood itself is unnormalized, in which case the model is often called a Gibbs distribution. For these, $\ell(\theta;x) \propto \tilde{\ell}(\theta,x)$, where $\tilde{\ell}(\theta,x)$ can be pointwise evaluated. Once again, working with $\nabla_x \log \ell(\theta;x)$ may be practical even when the normalizing constant is an intractable integral. Furthermore, when the likelihood can be written as the density of a natural exponential family model, $\nabla_x \log \ell(\theta;x)$ becomes linear in $\theta$, which is particularly useful in the development of new statistical methodology based on the Langevin Stein operator (see \cite{Barp2019}, 
\cite{Matsubara2021}).

\section{Computable Stein Discrepancies}\label{sec:comp_stein_discrepancies}


As mentioned in Section \ref{sec:stein-oper-discr}, many authors use Stein's method to assess IPMs between a target $P$ and some approximating measure $Q$ by  using Stein discrepancies  computed over sets $\mathcal{G}$ consisting of solutions to Stein equations.  In this section 
we will now show how 
Stein discrepancies may sometimes be computed \emph{exactly}  through 
a particular 
choice of Stein set 
(this issue was in fact already identified by \citet{stein2004use}). Exact computation turns out to be 
possible when
comparing an empirical measure $Q_n = n^{-1} \sum_{i=1}^n \delta_{x_i}$, with data points $x_i \in \X$, $\delta_{x_i}$ giving all probability mass to $x_i$,  to a given target distribution $P$. We will call any such discrepancy a \emph{computable Stein discrepancy}. 


The most common choice of computable discrepancies are the so-called \emph{kernel Stein discrepancies} (KSD), which use the unit-ball of a reproducing kernel Hilbert space (RKHS) as a Stein set, and can therefore be considered the Stein discrepancy counterpart to the \textit{maximum mean discrepancy} IPM \citep{gretton2006kernel,Smola2007,gretton2012kernel}.
An RKHS $\H_k$ is a Hilbert space (with norm $\|.\|_k$ and inner product $\langle \cdot,\cdot \rangle_k$) with an associated function $k:\X \times \X \rightarrow \R$ satisfying (i) symmetry; $k(x,y) = k(y,x)$ for all $x,y \in \mathcal{X}$, (ii) positive definiteness;  $\sum_{i,j=1}^n c_i c_j k(x_i, x_j) \geq 0$ for all $c_i \in \reals$, $x_i \in \X$, and (iii) the reproducing property $f(x) = \langle k(x,\cdot),f\rangle_k$ for all $f \in \H_k$, $x \in \mathcal{X}$. 
The function $k$ is called a \emph{reproducing kernel} \citep{aronszajn1950theory,schwartz1964sous}.
This choice of Stein set was inspired by the zero mean reproducing kernel theory of \citet{Oates2017}, 
used 
in \citet{Chwialkowski2016,liu2016kernelizedICML,Gorham2017} and extended in \citet{Sun2021}
to the case of matrix-valued kernels.
The main advantage is that the supremum in \eqref{eq:SteinDisc} can be analytically computed in terms of the reproducing kernel: 

\begin{example}[Langevin Kernel Stein discrepancies]\label{example:Langevin_KSD} 
The \emph{Langevin KSD} on $\X = \reals^d$ is obtained by combining the Langevin Stein operator $\operator{}$ from Example \ref{ex:steinopdiffusions} with a \emph{kernel Stein set} $\ksteinset{k} \defeq \{ g = (g_1,\dots, g_d) \mid \twonorm{v}\le 1 \text{ for } v_j \defeq \knorm{g_j} \}$:
\begin{talign}\label{ksd}
\ksd(Q) 
 & \defeq \mathcal{S}(Q,\operator{},\mathcal{G}_k)
 = \sqrt{\E_{X,X' \sim Q}[k_P(X, X')]} ,
\end{talign}
where the \emph{Stein reproducing kernel} is given by
\begin{talign}
  k_P(x,x') := & \operatorname{Trace}(\mathcal{T}_{x} \mathcal{T}_{x'} k(x,x') )  =  \langle \nabla_x , \nabla_{x'} k(x,x') \rangle + \langle \nabla_x k(x,x') , \nabla_{x'} \log p(x') \rangle \nonumber \\
    &  + \langle \nabla_{x'} k(x,x') , \nabla_x \log p(x) \rangle + k(x,x') \langle \nabla_x \log p(x) , \nabla_{x'} \log p(x') \rangle\label{eq:steinreproducingkernel}.
\end{talign}
Here the subscript in $\mathcal{T}_x$ indicates that the input of $\mathcal{T}$ is seen as a function of $x$. Most notably, this Stein reproducing kernel satisfies $\E_{X \sim P}[k_P(X,x)] = 0$ for all $x \in \mathbb{R}^d$ 
under mild regularity conditions (see \citet{Oates2017}). Whenever the approximating measure is $Q_n = n^{-1} \sum_{i=1}^n \delta_{x_i}$, the Langevin KSD has the simple closed form
\begin{talign}
\ksd(Q_n) = \mathcal{S}(Q_n,\operator{},\mathcal{G}_k) = \sqrt{ \frac{1}{n^2} \sum_{i,j=1}^n k_P(x_i,x_j) }.
\label{eq:KSD eqn}
\end{talign}
\end{example}
The most common choice of kernel $k$ is the inverse multi-quadric kernel $k(x,y) = (c^2 + \|x-y\|_2^2)^\beta$, $c > 0$, $\beta \in (-1,0)$. This is because \citet[][Theorem 8]{Gorham2017} showed that, if $\nabla \log p$ is sufficiently regular, then $Q_n$ converges weakly to $P$ whenever $\textsc{KSD}_k(Q_n) \rightarrow 0$. 
We will return to the implications of this in Section \ref{sec:measuring}.

Extensions of the Langevin KSD include
\citet{Barp2019}, who used the infinitesimal generator of general It\^{o} diffusions to get a family of \emph{diffusion kernel Stein discrepancies}; 
\citet{yang2018goodness} 
to discrete sets $\X$;
\citet{barp2018riemannian} 
to the case where $\X$ is a Riemannian manifold, such as in directional statistics.   


A second type of computational Stein discrepancies are the \emph{graph Stein discrepancies} (GSDs) of \citet{GorhamMa15,GorhamDuVoMa19}.

\begin{example}[Graph Stein discrepancies]
\label{ex:gsd}
The \emph{graph Stein discrepancies} combine a diffusion Stein operator $\operator{}$ as in  \cref{eqn:diffusion-stein-operator} with a \emph{graph Stein set}
\begin{talign*}
     \gsteinset{}{E} = \Big\{  g  &:
        \max\left(\|g(v)\|_\infty, \|\nabla g(v)\|_\infty,
        {\textstyle\frac{\|g(x) - g(y)\|_\infty}{\|x - y\|_1}},
    {\textstyle \frac{\|\nabla g(x) - \nabla g(y)\|_\infty}{\|x - y\|_1}}\right) \le 1,  \\
    &  \textstyle\frac{\|g(x) - g(y) - {\nabla g(x)}{(x - y)}\|_\infty}{\frac{1}{2}\|x - y\|_1^2} \leq 1,
    \textstyle\frac{\|g(x) - g(y) -{\nabla g(y)}{(x -
    y)}\|_\infty}{\frac{1}{2}\|x - y\|_1^2} \leq 1,  \\
    & \; \forall (x,y)\in E, v \in \supp{Q_n} \Big\}
\end{talign*}
where $\nabla g$ denotes the Jacobian matrix of $g$ and $E$ is a set of pairs of the form $(x_i,x_j)$, which must be taken sufficiently large to ensure that the GSD has Wasserstein convergence control \citep[][Theorem 2, Proposition 5 and 6]{GorhamMa15}. 
\end{example}
Once again, the Stein set is selected so that the discrepancy can be computed efficiently. 
The GSD is actually the solution of a finite-dimensional linear program, with the size of $E$ as low as linear in $n$, implying that it can be efficiently computed. 

While computable, both KSDs and GSDs suffer from a computational cost that grows at least quadratically in the sample size $n$. There exist at least two practical options for large sample sizes.  The \emph{finite set Stein discrepancies} of \citet{Jitkrittum2017} achieve a linear runtime by learning a small number of adaptive features based on Stein-transformed kernels,  so as to distinguish $P$ from  $Q$ samples with maximum test power. The \emph{random feature Stein discrepancies} of \citet{HugginsMa2018} approximate a broad class of convergence-determining Stein discrepancies in near-linear time using importance sampling.  To reduce the computational cost 
of Stein discrepancies 
in high dimensions,
the \emph{sliced Stein discrepancies }of \citet{Gong2020} can be used.

Finally, the computation of a Stein discrepancy can also be prohibitive if the Stein operator is expensive to evaluate. This commonly occurs in Bayesian and probabilistic inference where $\operator{} = \sum_{l=1}^L \opsub{l}$ is a sum over likelihood terms or potentials which are each more easily evaluated than $\mathcal{T}$ itself. To address this deficiency, \citet{gorham2020stochastic} introduced \emph{stochastic Stein discrepancies} (SSDs)
 \begin{talign}\label{ssd_subset}
    \ssd{Q_n} 
        &\defeq \sup_{g\in\gset}\left |\frac{L}{n}\sum_{i=1}^n
   \opsubarg{\sigma_i}{g}{x_i}\right | \qtext{for}
    \sigma_i \distiid \Unif(\{1,\dots, L\}).
\end{talign}
They showed that SSDs inherit the convergence control properties of standard discrepancies with probability 1. In \cite{xu2021stein}, for a special case of a stochastic Stein discrepancy, Stein's method is used to establish its asymptotic normality.

\section{New Statistical Methods for Assessing Sample Quality, Constructing Sample Approximations, and Improving Monte Carlo Integration}\label{sec:B}

This section 
details how ingredients from Stein's method have been successfully used to uncover methodological tools and procedures, and discusses a range of recent applications of Stein's method in computational statistics and machine learning. \cref{sec:measuring} shows how computable Stein discrepancies can be employed to quantify the quality of approximate MCMC schemes. \cref{sec: sampling sec} introduces a variety of ways of using Stein's method to  construct and improve a sample approximation, including Stein variational gradient descent (\cref{sec: SVGD}), Stein points (\cref{sec:steinpoints}), and Stein thinning (\cref{sec:steinthinning}). \cref{sec: CVs} describes Stein-based control variates for improved Monte Carlo integration, \cref{sec:estimators} presents statistical estimators, and 
\cref{sec:goodness-of-fit} details goodness-of-fit tests.

\subsection{Measuring Sample Quality} 
\label{sec:measuring}

This section presents practical tools based on Stein's method for computing how well a given sample, {represented as {an} empirical measure} $Q_n = n^{-1}\sum_{i=1}^n \delta_{x_i}$, approximates a given target distribution $P$.
This line of work was motivated by the approximate Markov chain Monte Carlo (MCMC) revolution in which practitioners have turned to asymptotically biased MCMC procedures that sacrifice asymptotic correctness for improved sampling speed \citep[see, e.g.,][]{WellingTe11,Ahn2012,Korattikara2014}. The reasoning is sound -- the reduction in Monte Carlo variance from faster sampling can outweigh the bias introduced, but standard Monte Carlo diagnostics like effective sample size, asymptotic variance, trace and mean plots, and pooled and within-chain variance diagnostics presume eventual convergence to the target distribution and hence do not account for asymptotic bias. To address this deficiency, \citet{GorhamMa15,Gorham2017,HugginsMa2018,gorham2020stochastic} introduced the computable Stein discrepancies of \cref{sec:comp_stein_discrepancies} as measures of sample quality suitable for comparing asymptotically exact, asymptotically biased, and even deterministic sample sequences $\{x_1, \dots, x_n\}$.

\vspace{-3mm}

\begin{figure}
    \centering
   \includegraphics[width=0.3\textwidth]{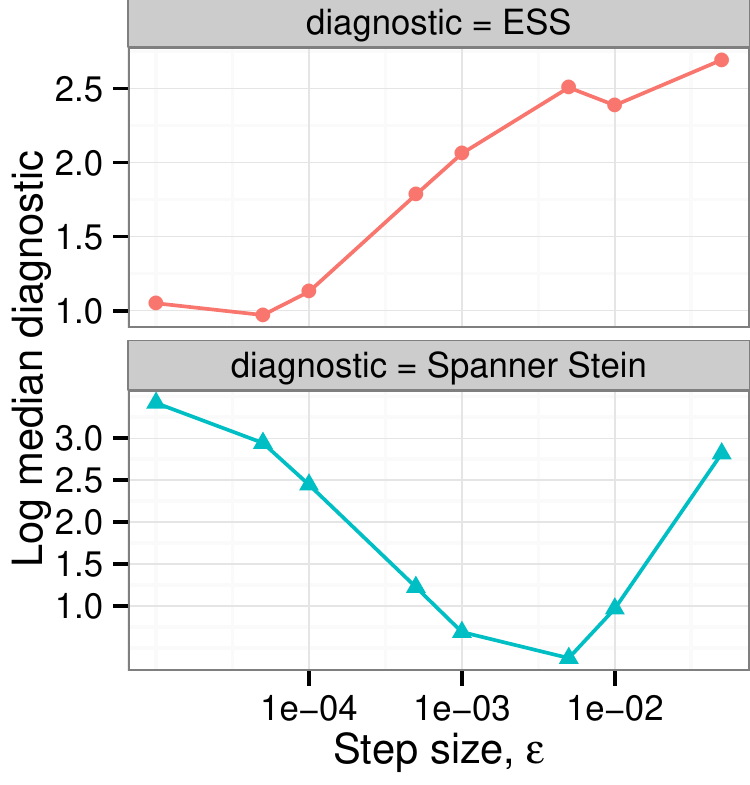}
   \includegraphics[width=0.67\textwidth]{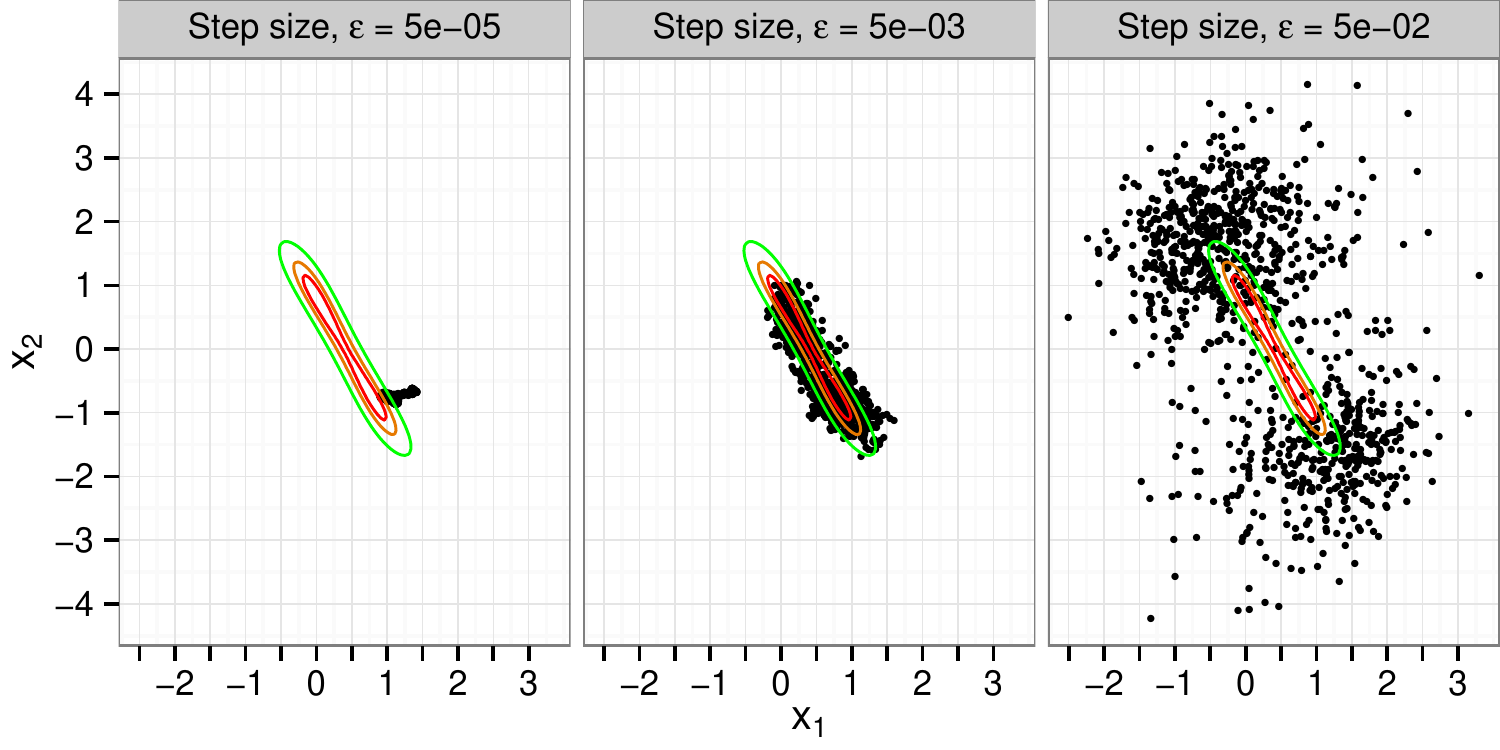}
   \caption{Selecting the step size $\eps$ for stochastic gradient Langevin dynamics \citep{WellingTe11}, a popular approximate MCMC algorithm designed for scalability. Standard MCMC diagnostics like effective sample size (ESS) do not account for asymptotic bias and select overly large $\eps$ with greatly overdispersed samples (right panel). Overly small $\eps$ leads to slow mixing (left panel).
   The Stein discrepancy selects an intermediate value offering the best approximation (center panel). Figure reproduced from \citet[Fig.~3]{GorhamMa15}.}
    \label{fig:gsd-ess-sgld}
\end{figure}

\paragraph{Graph Stein discrepancies}
\citet{GorhamMa15} used the GSDs of \cref{ex:gsd} to select and tune approximate MCMC samplers, assess the empirical convergence rates of Monte Carlo and Quasi-Monte Carlo procedures, and quantify bias-variance tradeoffs in posterior inference.
An illustrative example is given in \cref{fig:gsd-ess-sgld}.  
These applications were enabled by a series of analyses establishing that the GSD converges to $0$ if and only if its empirical measure $Q_n$ converges to $P$.
Specifically, \citet{mackey2016multivariate,GorhamDuVoMa19,ErdogduMaSh2018} bounded the GSD explicitly above and below by Wasserstein distances whenever the diffusion underlying the Stein operator couples quickly and has pseudo-Lipschitz drift. 

\vspace{-3mm}

\paragraph{Kernel Stein discrepancies}
The closed form of the KSDs of \cref{example:Langevin_KSD} represents a significant practical advantage for sample quality measurement, as no linear program solvers are necessary, and the computation of the discrepancy can be easily parallelized. However, \citet{Gorham2017} showed that not all KSDs are suitable for measuring sample quality. In particular, in dimension $d \geq 3$, KSDs based on popular kernels like the Gaussian and \Matern kernels fail to detect when a sample is not converging to the target, even when the target is normal. To address this shortcoming, \citet{Gorham2017} developed a theory of weak convergence control for KSDs and designed a class of KSDs that provably control weak convergence for a large set of target distributions (see \citet{HugginsMa2018,Chen2018} for further developments). These convergence-determining KSDs have been shown to deliver substantial speed-ups over the original GSDs in higher dimensions \citep{Gorham2017}.

\vspace{-3mm}

\newcommand{\RFSDs}{$\textup{R}{\Phi}\textup{SDs}$\xspace}
\paragraph{Random feature Stein discrepancies} To identify a family of convergence-determining discrepancy measures that can be accurately and inexpensively approximated with random sampling, \citet{HugginsMa2018}  introduce a new domain for the Stein operator using a feature function, giving rise to a {\it feature Stein set} and a corresponding {\it feature Stein discrepancy}. The feature Stein discrepancy is then approximated using importance sampling, which results a {\it{random feature Stein discrepancy}} (R$\Phi$SD). \citet{HugginsMa2018} showed that \RFSDs upper bound standard discrepancy measures with high probability. This translates into high-probability convergence control whenever the approximating sample sequence is uniformly integrable.

\renewcommand{\ff}{g}
\renewcommand{\ffs}{g^*}
\renewcommand{\F}{\mathcal{G}}
\renewcommand{\top}{\intercal}
\newcommand{\sumsteinp}{\sumstein}

\subsection{Constructing and Improving Sample Approximation}
\label{sec: sampling sec}
{Popular stochastic Monte Carlo methods such as MCMC provide a standard approach for constructing and improving a sample-based approximation of the form  $Q_n = n^{-1} \sum_{i=1}^n \delta_{x_i}$ for an intractable distribution $P$ of interest.}
In this section, we {explain how} 
Stein's method  
{can be used}
to develop a suit of \emph{optimization-based} alternatives to Monte Carlo methods. We demonstrate this with three examples:  \cref{sec: SVGD} introduces Stein variational gradient descent, a gradient based algorithm that iteratively updates the location of the particles $\{x_1, \ldots, x_n\}$ to improve the approximation quality w.r.t $P$. \cref{sec:steinpoints} introduces  Stein Points, a greedy algorithm that constructs the approximation by sequentially adding the particles to minimize KSD. 
\cref{sec:steinthinning} introduces Stein Thinning, which compresses an existing approximation using KSD.

\subsubsection{Sampling with Stein Variational Gradient}
\label{sec: SVGD}

Let $P$ be a distribution with a  continuously differentiable density function $p$ supported on $\X$. We want to find a set of  points $\{x_1, \ldots, x_n\}\subset \X$, which we refer to as \emph{particles}, such that its empirical measure $Q$ gives a close approximation to $P$. Stein variational gradient descent (SVGD) \citep{Liu2016a} achieves this by iteratively updating the particles to minimize the KL divergence between $Q$ and $P$, which is made possible by exploiting an intrinsic connection between KL divergence and Stein's method, as follows.

For the purpose of derivation, we assume for now that $Q$ is a continuous distribution with a finite KL divergence $\KL(Q~||~P) <\infty$. We want to recursively ``transport" the probability mass of $Q$ with a deterministic map to move it closer to $P$ in order to decrease $\KL(Q~||~P)$ as fast as possible. Specifically, we consider mappings  of the form 
\begin{talign*}
\T(x) = x + \epsilon \ff(x), 
\end{talign*}
where $\epsilon$ is a small positive scalar that serves as a step size, and 
$\ff\colon \X \to \X$ 
is a one-to-one mapping that serves as the velocity field. 
Denote by $\Ts \muo$ 
the  distribution of $\T(X)$ when $X\sim \muo$; this is also called the {\it pushforward measure}. 

\begin{figure}[t!]
\centering
\includegraphics[width = 0.3\textwidth]{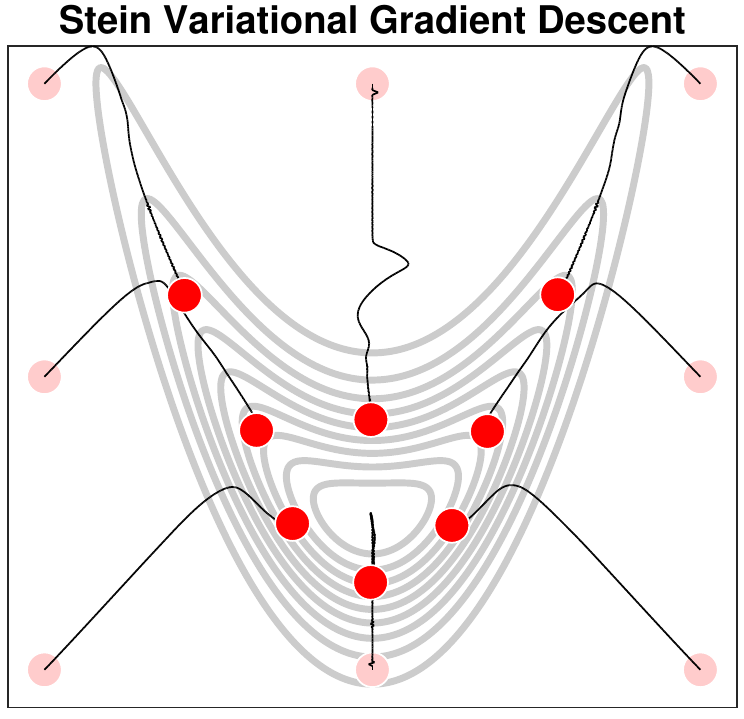} 
\hspace{5pt}
\includegraphics[width = 0.3\textwidth]{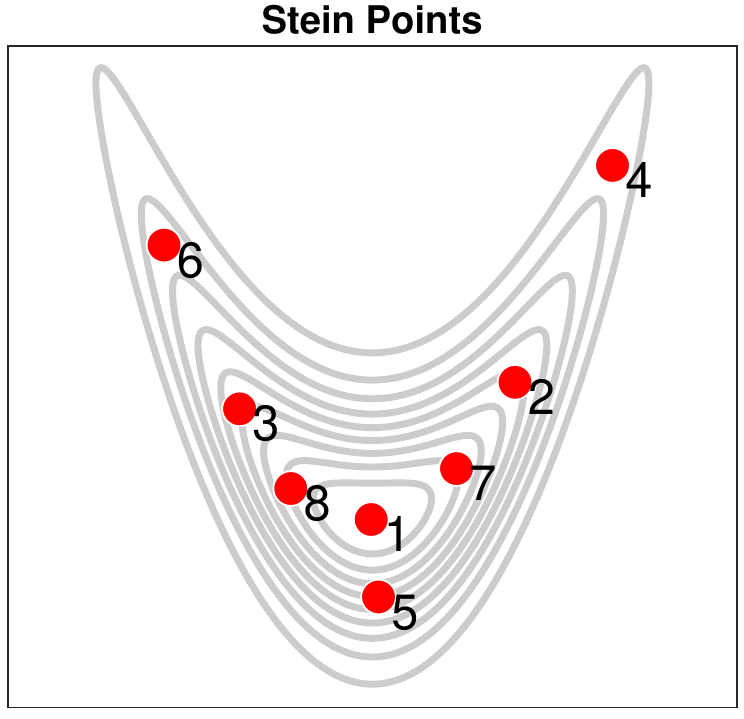} \hspace{5pt}
\includegraphics[width = 0.3\textwidth]{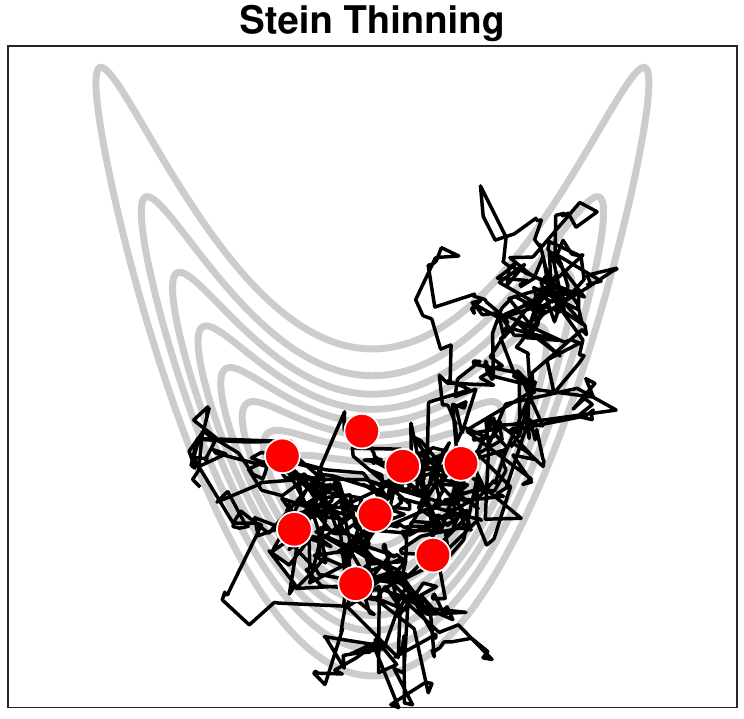}
\caption{Sampling with Stein's method. 
Left: The initial (transparent red) and final (red) states of 8 particles, together with their trajectories (black) under the \textit{Stein variational gradient descent} algorithm in \eqref{equ:update11}.
Middle: The first 8 states (red) of an extensible sequence produced by the \textit{Stein points} algorithm in \eqref{alg: stein points}. The order in which the states are selected is indicated.
Right: The first 8 representative states (red) selected from a Markov chain sample path (black), according to \eqref{alg: stein points 2}.
(Grey contours are shown for the distributional target, which in each case the red states are intended to represent.)
}
\label{fig:sp-mcmc}
\end{figure}

The key challenge is to optimally choose $\ff$ for each given $Q$, so that the KL divergence between $\Ts Q$ and $P$ is decreased as much as possible. Assuming $\epsilon$ is infinitesimal, the optimal choice of $\ff$ can be framed into a  functional optimization problem: 
\begin{talign} \label{equ:ff00}
\max_{\ff \in \F}  \left\{-\frac{\dno}{\dno\epsilon} \KL( \Ts \muo  ~|| ~ P) ~ \big |_{\epsilon = 0}  
\right\}, 
\end{talign}
where the negative derivative 
$-\frac{\dno}{\dno\epsilon} \KL( \Ts \muo  ~|| ~ P) ~ \big |_{\epsilon = 0}$ measures the decreasing rate of KL divergence under the transport map $\T$ as we increase the step size $\epsilon$ starting from zero, and  
 $\F$ is a function space that specifies the candidate set of $\ff$. The key observation is that the objective in \eqref{equ:ff00} is in fact equivalent to the expectation $\E_{Q}[\oparg{g}{X}]$ of the Langevin Stein operator. 
\begin{theorem}\label{thm:svgd_kl} 
Assume $P$ and $Q$ have positive densities on $\X=\RR^d$, and the density $p$ of $P$ is in $C^1(\RR^d)$. Let $\T(x) = x + \epsilon \ff(x)$, where $\epsilon \in \R$ and $\ff\colon \RR^d \to \RR^d$ is a $C^1$ map with $\sup_{x\in   \RR^d}\norm{\nabla\ff(x)}_2 <\infty$, where $\norm{\cdot}_2$ denotes the spectral norm. We have 
\begin{talign*} 
-\frac{\dno }{\dno \epsilon} \KL(\Ts Q ~||~  P)~\big|_{\epsilon=0} =  \E_{X \sim Q}[\oparg{g}{X}] \end{talign*}
where $
\oparg{g}{x} = 
\langle \nabla \log p(x),  \ff(x)\rangle  + \langle \nabla, \ff(x)\rangle$. 
\end{theorem}
Theorem~\ref{thm:svgd_kl} draws an intriguing connection between Stein's method, the KL divergence and optimal transport. 
It shows that \eqref{equ:ff00} is equivalent to the optimization in Langevin KSD:
\begin{talign}\label{equ:tpphi}
\mathrm{KSD}_k(Q)
= \max_{g \in \F} \left \{\E_{X \sim Q}[{\oparg{g}{X}}] \right \} 
= 
\max_{\ff \in \F}  \left \{ -\frac{\dno }{\dno \epsilon} \KL(\Ts Q ~||~  P)\big|_{\epsilon=0}  \right \}. 
\end{talign}
Therefore, the Langevin KSD can be interpreted as the maximum decreasing rate of KL divergence between $Q$ and $P$
under the best transport map in $\F$. Taking $\mathcal{G}$ to be the unit ball of the RKHS with kernel $k$, we can solve equation \eqref{equ:tpphi} in closed form (see Example~\ref{example:Langevin_KSD}):  
\begin{talign}\label{equ:ffs}
\ffs_{\muo,P}(\cdot) \propto  \E_{X \sim \muo}\left[
\nabla \log p(X) k(X,\cdot) + \nabla_x k(X,\cdot)\right]. 
\end{talign}
This yields the best update direction for ``transporting'' particles from $Q$ to $P$ under KL divergence.  In practice, we take $Q = n^{-1} \sum_{i=1}^n  \delta_{x_i}$ to be the empirical measure of the particles
while iteratively updating $\{x_1,\ldots,x_n\}$ by using the optimal transport map found above{,} $\T^*_{Q,P}(x) = x + \epsilon \ffs_{Q,P}(x)$. This yields the following simple update rule on the particles, which is illustrated in the left panel of \Cref{fig:sp-mcmc}: 
\begin{talign}
\label{equ:update11}
& x_i ~ \gets ~ x_i  ~  + ~ \epsilon \frac{1}{n}\sum_{j=1}^n \left ( \nablax \log p(x_j) k( x_j, x_i) + \nablax_{x_j} k(x_j, x_i) \right), ~~~ \forall i =1,\ldots, n.  
\end{talign}
The two terms in \eqref{equ:update11} play intuitive roles. The term with the gradient $\nabla \log p$ pushes the particles towards the high probability regions of $P$, while the term with $\nabla_x k$ can be viewed as a repulsive force to enforce  the diversity between the particles if $k$ is a stationary kernel of form $k(x,x') = \phi(x-x')$: in this case, performing  $x_i'  \gets x_i + \epsilon \nabla_{x_j} k(x_j, x_i)$ would decrease $k(x_i,x_j)$,  which measures the similarity between $x_i$ and $x_j$, when $\epsilon$ is sufficiently small. If there is no repulsive force, 
or when there is only a single particle (and the kernel satisfies $\nabla_x k(x,x') = 0$ for $x = x'$), the solution would collapse to the local optima of $\log p$, reducing to the maximum a posteriori (MAP) point. Therefore, by using different particle sizes,  SVGD provides an interpolation between  MAP to a full particle-based approximation. 

SVGD defines a \emph{deterministic interacting particle}  system in which $\{x_1, \ldots, x_n\}$ interact and co-evolve to reach a desirable equilibrium. 
For understanding SVGD asymptotically,
\citep{liu2017stein}
considers the limit of large particle size ($n\to\infty$) and continuous time ($\epsilon\to0$),  and  interprets SVGD as a \emph{gradient flow} of KL divergence induced by a kernel-Wasserstein geometric structure on the infinite dimensional space of distributions; a set of theoretical studies along this line can be found in \citet{lu2018scaling, liu2019understanding,duncan2019geometry, chewi2020svgd, gorham2020stochastic,nusken2021stein,korba2020non}. 
In the non-asymptotic regime of a finite number $n$ of particles,
SVGD acts like a \emph{numerical quadrature} method in which the particles are arranged to exactly estimate the true expectation of a set of special basis functions determined by the Stein operator and kernel function \citep{liu2018stein}.

SVGD has been extended and improved in various ways. For example, amortized SVGD \citep{feng2017learning} learns neural samplers in replacement of particle approximation; gradient-free SVGD \citep{han2018stein} provides an extension that requires no gradient information of the target distribution $P$; a number of other extensions and improvements can be found in, e.g., \citet{wang2017stein, zhuo2018message,wang2019stein,liu2017riemannian,detommaso2018stein,chen2018unified,chen2019projected,li2020stochastic,gorham2020stochastic,gong2019quantile,wang2019nonlinear,han2017stein}. SVGD {has  found} applications in a variety of problems including in deep learning \citep[e.g.,][]{ pu2017vae, wang2016learning}, reinforcement learning \citep[e.g.,][]{haarnoja2017reinforcement,liu2017policy, liu2020off}, meta learning \citep[e.g.,][]{kim2018bayesian,feng2017learning}, and uncertainty quantification in  science and engineering \citep[e.g.,][]{zhu2018bayesian, zhang2019seismic, zhang2020variational, zhang2019bayesian}.

\subsubsection{Sampling with Stein Points}\label{sec:steinpoints}
The \emph{Stein points} \citep{Chen2018,Chen2019}  approach 
progressively constructs a set of points  $\{x_1,\ldots,x_n\} \subset \X$ to approximate $P$ by minimizing a Stein discrepancy. 
For example, the KSD 
can be minimized in a sequential greedy manner: $x_1  \in  {\mathrm{argmin}}_{{x \in \mathcal{X}}} \text{KSD}_k(\{x\})$ and 
\begin{talign}
x_n  \in  \mathrm{argmin}_{x \in \mathcal{X}} \text{KSD}_k(\{x_1,\dots,x_{n-1},x\}) \: \text{for } n > 1, \label{alg: stein points}
\end{talign}
where $\text{KSD}_k(\{x_1, \ldots, x_n\}) = \mathcal{S}(Q_n,\operator{},\mathcal{G}_k)$ and 
$\mathcal{G}_k$ is a kernel Stein set;
then the  set $\{x_1, \ldots, x_n\}$ is selected
as to approximately minimize this KSD. A typical sequence obtained in this way is presented in the middle panel of \Cref{fig:sp-mcmc}.

Finding the global minima in  equation \eqref{alg: stein points} 
may be difficult. However,
\citet[][Theorem 2]{Chen2018} showed that even imperfect optimization methods can lead to a fast decrease of the KSD. More precisely, if $k_P$ in equation \eqref{eq:steinreproducingkernel} satisfies $\mathbb{P}_{X\sim P}( k_P(X,X) \geq t ) \leq b_1 \mathrm{e}^{-b_2t}$ for some constants $b_1, b_2 > 0$ and all $t \geq 0$,  then there exist constants $c_1, c_2 > 0$ depending only on $k_P$ and $P$ such that any  $n \in \mathbb{N}$ and $\{x_1, \ldots, x_n \} \subset \mathcal{X}$ satisfying 
\begin{talign*}
\textsc{KSD}_k(\{x_1,\dots,x_j\})^2
 \leq 
\frac{\delta}{n^2} + \min_{x\in \mathcal{X}: k_P(x,x)\leq \frac{2\log(j)}{c_2}} \textsc{KSD}_k(\{x_1,\dots,x_{j-1},x\})^2 
\end{talign*}
for all $j=1, \ldots, n,$ lead to an upper bound on the KSD of the form
\begin{talign*}
\textsc{KSD}_k(\{x_1, \ldots,x_n\}) 
	 \leq \mathrm{e}^{\pi/2} \sqrt{\frac{2\log(n)}{c_2n} +\frac{c_1}{n} + \frac{\delta}{n}}.
\end{talign*}
Thus, KSD can be used to transform the sampling problem of approximating $P$ into an optimization problem that admits a provably convergent numerical method.


\subsubsection{Stein Thinning}\label{sec:steinthinning}

\cite{riabiz2020optimal} 
use 
KSD in a post-processing approach to select states from a large pre-determined candidate set, with application to de-biasing MCMC output. Their approach can be summarized as:
\begin{talign}
x_1  & \in  \mathrm{argmin}_{x \in \{X_1,\dots,X_N\}} \text{KSD}_k(\{x\}), \nonumber \\
 x_n  & \in  \mathrm{argmin}_{x \in \{X_1,\dots,X_N\}} \text{KSD}_k(\{x_1,\dots,x_{n-1},x\})\: \text{for } n > 1, 
\label{alg: stein points 2}
\end{talign}
where $(X_i)_{i =1, \ldots, N}$ is a $Q$-invariant Markov chain; $Q$ and $P$ need not be equal.
A typical sequence obtained in this way is presented in the right panel of \Cref{fig:sp-mcmc}.
These authors extended earlier convergence results to prove almost sure weak convergence of $Q_n = n^{-1} \sum_{i=1}^n \delta_{x_i}$ to $P$ in the limit as $N \geq n \rightarrow \infty$.
Indeed, provided that the Markov chain is $V$-uniformly ergodic with $V(x) \geq \frac{\mathrm{d}P}{\mathrm{d}Q}(x) \sqrt{k_P(x,x)}$ and that certain moments of the chain are finite,  \citet[][Theorem 3]{riabiz2020optimal} showed that $\text{KSD}_k(\{x_1, \ldots, x_n\}) \rightarrow 0$ almost surely as $n \rightarrow \infty$.

Thus, Stein discrepancies may be used to post-process MCMC output, which
can have the benefits of improving approximation quality, mitigating sampler bias, and providing a compressed representation of $P$. The closed form of KSD renders such post-processing straightforward. Extensions of Stein thinning, to allow for non-myopic optimization and for mini-batching, were recently studied in \cite{teymur2020optimal}. In related work, \cite{liu2016black,hodgkinson2020reproducing} proposed to use Stein discrepancies to re-weight Markov chain output, as opposed to selecting a smaller subset.

\subsection{Improving Monte Carlo Integration}
\label{sec: CVs}

{As already mentioned,} the  problem of approximating expectations $\mathbb{E}_{X \sim P}[ f(X) ]$, where $f:\X \rightarrow \R$ is a test function of interest, is at the heart of 
Stein's method, see \cite{stein1986approximate}. In Bayesian statistics, it is most common for expectations to be approximated using ergodic averages from MCMC, though of course the algorithms described in Sections \ref{sec: SVGD} and \ref{sec: sampling sec} can also be used. The convergence of estimators based on MCMC is characterized by the central limit theorem, whose asymptotic variance will depend on the variance of $f$ along the sample path of the Markov chain \citep[see Chapter 17 of][]{meyn2012markov}. In \cite{stein2004use}, auxiliary variables are constructed for such variance reduction in  a particular setting. A recent approach to reducing the asymptotic variance is to use so-called \emph{control variates}. This consists of designing a function $h:\X \rightarrow \R$ such that, if we re-write the expectation as
\begin{talign*}
\mathbb{E}_{X \sim P}[f(X)] = \mathbb{E}_{X \sim P}[h(X)] + \mathbb{E}_{X \sim P}[f(X) - h(X)],
\end{talign*}
then the first term on the right-hand side is known analytically (by some auxiliary argument) and the second integrand, $f - h$, should have smaller variance than $f$ along the sample path of the Markov chain.
In this way estimation of the original expectation is reduced to estimation of an alternative expectation which is more amenable to MCMC.
Indeed, in an ideal situation we would pick $h$ such that $f-h$ is constant along the sample path of the Markov chain, 
so that the ergodic average is exact after just one iteration of the chain has been performed
\citep{Mira2013}.

The principal limitation to the successful application of control variates is the identification of a set of candidates for $h$ that (a) is sufficiently rich to approximate $f$ and (b) for which the expectations $\mathbb{E}_{X \sim P}[h(X)]$ can be evaluated. Several authors have developed bespoke solutions that are specific to a particular MCMC algorithm, including \citet{Andradottir1993,stein2004use,henderson2004adaptive,dellaportas2012control,mijatovic2018poisson}. It was pointed out in \cite{Oates2017} that the image of a Stein operator adapted to $P$ can serve as such a set in general. 
In concrete terms, one may identify a Stein operator $\operator{}$ and a Stein set $\mathcal{G}$ that are adapted to $P$ and then attempt to pick an element $g \in \mathcal{G}$ for which $f - h \approx \text{constant}$ along the Markov chain sample path, where $h = \operator{g}$. This problem is closely related to numerical solution of the Stein equation \eqref{eq:steinequ}. 

In \cite{Assaraf1999,Mira2013,Oates2016thermo}, the authors selected $g$ from the set of all polynomials of a fixed maximum degree, minimizing the squared error $J_n(g) = \sum_{i=1}^n (f(x_i) - \mathcal{T}g(x_i))^2$ along the Markov chain sample path $\{x_1,\ldots,x_n\}$, with no complexity penalty used.
In \cite{south2018regularised}, the authors used an $\ell_1$ or $\ell_2$ penalty on the polynomial coefficients and recommended 
cross-validation
as a means to select an appropriate polynomial degree.
Kernel methods with a minimum norm penalty were proposed in \cite{Oates2017,Oates2016CF2,Barp2019,Sun2021}.
In \cite{south2020semi}, the authors showed how polynomials and reproducing kernels can be combined in a manner that leads to polynomial exactness of the control variate estimator in the Bernstein--von--Mises limit.
The use of neural networks for $g$ was empirically assessed in \cite{Zhu2018,Si2020}.
If one specializes to particular MCMC algorithms then it may be possible to consistently estimate the asymptotic variance under the Markov chain, which can be used to construct a more appropriate functional $J_n$.
This approach is exemplified in \cite{belomestny2017variance,belomestny2019variance,belomestny2020variance}.  
\cite{liu2017action} 
provides a detailed application of Stein control variates to policy optimization in reinforcement learning.

The diverse set of approaches for constructing control variates based on Stein operators supports the view that no single method will be universally optimal for all real-world computational problems and, to some extent, the estimation of a suitable control variate remains as much an ``art'' as the design of an efficient MCMC method.

\subsection{Statistical Estimators based on Stein discrepancies}\label{sec:estimators}

Computable Stein discrepancies have also been used for parameter estimation. Let $\mathscr{P}_\Theta=\{P_\vartheta:\,\vartheta\in\Theta\}$ denote a parametric family of distributions, and assume we would like to recover the element of this family which generated some data $x_1, \ldots, x_n$ (represented by $Q_n = n^{-1} \sum_{i=1}^n \delta_{x_i}$).  \citet{Barp2019} proposed \emph{minimum Stein discrepancy estimators}, which are a general class of estimators of the form
\begin{talign}\label{eq:minimumSD}
\hat{\vartheta}_n := \arginf_{\vartheta \in \Theta} \stein{Q_n}{\operator{}_{\vartheta}}{\gset},
\end{talign}
where $\operator{}_{\vartheta}$ is a Stein operator characterising $P_\vartheta$, and showed that a number of machine learning algorithms including score-matching \citep{hyvarinen2005estimation}, contrastive divergence \citep{Hinton2002}, and minimum probability flow \citep{Sohl-dickstein2011} are specific instances of this framework.
have also been proposed. \citet{Barp2019} studied the special case of minimum diffusion KSD estimators and showed that these enjoy desirable robustness properties under regularity conditions on the kernel. This was then studied further in the context of discrete models by \citet{Banerjee2021}, whilst \citet{grathwohl2020learning} considered a Stein discrepancy where the Stein space is indexed by a neural network.  Relatedly, \citet{betsch2020minimum} studied minimum $L^q$ distance estimators based on Stein operators, and \citet{Liu2018Fisher} considered a minimum distance estimator based on likelihood ratios estimated through Stein operators.

Most notably, estimators of the form in \eqref{eq:minimumSD} are useful for unnormalised likelihood models, since Stein operators usually rely on unnormalised densities. When the parametric family is in some exponential family, the Langevin Stein discrepancies become quadratic forms in $\vartheta$, which implies that the optimiser can be obtained in closed-form.
\citet{Matsubara2021} built on this idea to propose a fully conjugate generalised Bayesian approach for unnormalised densities. This was latter extended to discrete data settings by \citet{Matsubara2022}.

\section{New Methods for and Insights in Statistical Inference via Stein Operators and Stein Discrepancies}\label{sec:goodness-of-fit}

In this section we focus on statistical inference and show how tools from Stein's Method have been put to use to build new powerful tools as well as to gain novel insights in long-existing procedures. Section~\ref{kernelSteinEmbedding} is concerned with new goodness-of-fit tests obtained from Stein discrepancies, while Section~\ref{sec:stein-oper-goodn} deals with composite goodness-of-fit tests based on Stein operators. These goodness-of-fit tests lend themselves very naturally to a further Stein-based analysis, namely to quantify the distance, at a given finite sample size $n$, between
the asymptotic distribution and the unknown exact distribution, hereby getting an idea
of how good the asymptotic approximation actually is. 
More generally recently Stein’s method has been used to quantify the asymptotic behaviour
of statistical estimators and hypothesis tests, which is the topic of Section~\ref{sec:MLE}. In a similar vein, {Section~\ref{sec:Bayes}} deals with the Bayesian setting, hereby showing a new way to quantify the finite-sample effect of the prior choice.

\subsection{Goodness-of-fit Tests from Stein Discrepancies}\label{kernelSteinEmbedding}

Suppose we would like to test for the null hypothesis $H_0: Q = P$ based on realizations $\{x_1, \ldots, x_n\}$ from $Q$ (which may or may not be independent). 
\citet{Chwialkowski2016,Liu2016test} 
proposed to use a KSD as test statistic, which is particularly powerful for a distribution $P$ whose density is known up to a normalizing constant.

These tests are motivated by the general approach of using IPMs within a hypothesis testing framework. In particular, an influential line of work in machine learning has been to use IPMs with a kernel-based underlying function class, leading to the so-called MMD hypothesis tests \citep{gretton2006kernel,gretton2012kernel}.
This approach has previously been used to test for a range of hypotheses, including two-sample tests and independence tests. Their popularity can be explained through their generality: they only rely on the choice of a kernel and samples from
both $P$ and $Q$, and can hence be implemented for a wide range of problems.

In the goodness-of-fit setting, when $P$ has a density known up to normalizing, sampling from $P$ may introduce unnecessary variance to our test statistic. The test is also somewhat sub-optimal since it does not use any specific properties of $P$. It is therefore natural to consider the use of Stein operators in this setting. 
This can be achieved by selecting an IPM whose underlying function class is of the form $\operator{g}$ for $g$ in some Stein set $\mathcal{G}$. When using a Langevin Stein operator and kernel Stein set, this leads to the Langevin KSD of Example \ref{example:Langevin_KSD}, which is the case most often considered in this literature. Recalling the expression for the population Langevin KSD given in equation (\ref{ksd}),
an unbiased estimate of the squared KSD takes the
convenient form of a U-statistic:
\begin{talign*}
  \widehat{\mathrm{KSD}}_{k}^2(Q)=
  \frac{2}{n(n-1)}
  \sum_{i<j}
  k_P  (x_i,x_j).
\end{talign*}
This estimate can be used as a test statistic. It is degenerate under the null hypothesis that $Q=P,$ and non-degenerate
under the alternative. As a result, when the sample is i.i.d.\  the asymptotic behaviour of the
statistic is obtained via standard results \citep{serfling2009approximation}. Unfortunately, the asymptotic distribution under the null is a function of the eigenvalues of $k_P$  with respect to $Q$, which are rarely computable in closed form.
Nonetheless, a test threshold of asymptotic level $\alpha$ may be obtained using a wild-bootstrap procedure on a V-statistic approximation to the KSD. The wild bootstrap may also be adapted to the case where the sample from $Q$ is not i.i.d., but satisfies a $\tau$-mixing condition \citep{LeuNeu13}. This is especially helpful when the goodness-of-fit test is used for bias quantification of approximate MCMC procedures since these are not i.i.d.\ \cite[Section 4]{Chwialkowski2016}.

In order to guarantee consistency of the tests, it is of interest to establish when the KSD uniquely determines
whether $Q$ and $P$ correspond. We refer to \citet[Theorem 2.2]{Chwialkowski2016}:
if $k$ is $C_{0}$-universal \cite[Definition 4.1]{Carmeli2010},
and if $\mathbb{E}_{X\sim Q}[\Vert \nabla(\log (p(X)/q(X)))\Vert_2 ^{2}]<\infty$,
then $\mathrm{KSD}_{k}(q)=0$ if and only if $P=Q$. Many popular kernels,
including the exponentiated quadratic (Gaussian) kernel $k(x,y) = \exp(-\|x-y\|^2_2/l^2)$ $(l>0)$, are $C_{0}$-universal. We however recall the result of \citet{Gorham2017} that stronger conditions on the kernel are required when one wishes to control {\em weak  convergence} to a target using the KSD.

Apart from U-statistic based tests, alternative tests exist which
can be computed in linear time, using adaptive kernel
Stein features that indicate where the data distribution $Q$ differs
from the model $P$  \citep{Jitkrittum2017}, or importance sampling approaches \citep{HugginsMa2018}. In the former case, the features are learned on a  held-out sample from $Q$, so
as to maximize the power of the resulting test.

Stein goodness-of-fit
tests may also be defined for right-censored time-to-event data. Indeed, \citet{fernandez2020kernelized}
defined three  Stein operators for this setting, which exploit well-known identities in survival analysis that arise from the underlying  structure of the data. The first is the {\em Survival Stein Operator}, which arises from a direct application of the Langevin Stein operator to the density function; 
the second, the {\em Martingale Stein Operator}, applies a well-known Martingale equality in a similar fashion as for log-rank statistics; and the third,  the {\em Proportional Stein Operator}, applies the Langevin Stein operator to the  hazard function.
The resulting Stein tests were used 
to validate models of survival times
in  real-world medical studies of Leukemia, Chronic Granulotamous Disease, Ovarian Cancer, and Lung Cancer.



For discrete distributions, KSD tests include the work of \citet{yang2018goodness} which  derives a discrepancy for discrete data and that of \citet{yang2019stein} which focuses on point processes.
{For exponential random graph models when only one network observation is available,
\cite{xu2021stein}  use the Stein operator for exponential random graph models from \cite{reinert2019approximating} as basis for a kernelised Stein discrepancy test.}


\subsection{Composite Goodness-of-fit Tests from Stein Operators}
\label{sec:stein-oper-goodn}

Consider the classical problem of testing the composite {null hypothesis} $H_0:\;Q\in\mathscr{P}_\Theta=\{P_\vartheta:\,\vartheta\in\Theta\}$, where $\Theta\subset\R^s$, $s\in\N$, is an open parameter space, and $P_\vartheta$ is the unique distribution corresponding to $\vartheta\in\Theta$ in the parametric family $\mathscr{P}_\Theta$. 
 This hypothesis is to be tested based on an 
 i.i.d.\ sample $\{x_1, \ldots, x_n\}$ from $Q$. 
 For example tests for normality fall into this category. 
 
 For this problem,  test statistics based on 
 parametric families of Stein operators as in   \citet{ley2016}  have been developed as follows. 
 Let $\{\opsub{\vartheta}:\vartheta\in\Theta\}$ be a family of Stein operators characterizing the family $\mathscr{P}_\Theta$. By the Stein characterization we have $\mathbb{E}_{X \sim P_\vartheta}[(\opsub{\vartheta} g)(X)]=0$ for all $g\in\mathcal{G}(\opsub{\vartheta})$ and $\vartheta\in\Theta$.
 A natural extension of the KSD framework to the composite hypothesis was proposed by \cite{Key2021} and built on the minimum Stein discrepancy estimators of \cite{Barp2019}. However, we will focus this section on an alternative test for the composite hypothesis based on a suitable set of test functions $\mathcal{G}=\{g_t(x): t\in M\}$, $M\subset\R^d$  given by the weighted $L^2$ statistic
\begin{talign}\label{eq:Tn}
T_n
& = n \int_{M}\big\|\frac{1}{n}\sum_{i=1}^n(\opsub{\widehat{\vartheta}_n}g_t)(x_i) - \E_{X \sim P_{\widehat{\vartheta}_n}}[(\opsub{\widehat{\vartheta}_n}g_t)(X)]\big\|^2\omega(t)\,\mbox{d}t  \\
& = n\int_{M}\big\|\frac1n\sum_{i=1}^n(\opsub{\widehat{\vartheta}_n}g_t)(x_i)\big\|^2\omega(t)\,\mbox{d}t,\nonumber
\end{talign}
where $\widehat{\vartheta}_n$ is a consistent estimator of $\vartheta$, $\|\cdot\|$ is a suitable norm and $\omega:M\rightarrow[0,\infty)$ is a positive weight function satisfying some weak integrability conditions. Heuristically, $T_n$ should be close to 0 if and only if the data stems from $\mathscr{P}_\Theta$, and we 
hence reject $H_0$ for large values of $T_n$.

\citet{Henze2020} implicitly used such a test for multivariate normality based on the classical Stein operator $\operator{}$ from Example \ref{ex:steingauss}.
An alternative test of univariate normality based on $\operator{}$ from Example \ref{ex:steingauss} is  proposed in \citet{ebner2020combining}, but in this case test functions of the form $\{g_t(x)=\exp(\mathrm{i}t x): t\in\R\}$ (i.e. related to characteristic functions) are used. \citet{Doerr2020} also introduce a test of multivariate normality, based on  $\oparg{g}{x}=-\Delta g(x)+(\|x\|_2^2-d)g(x)$ (where $\Delta$ denotes the Laplacian), and the class of test functions $\{g_t(x)=\exp(\mathrm{i}t^\top x): t\in\R^d\}$. There are considerable differences in power against specific alternatives between the tests, especially w.r.t.\ the choice of test functions. For a comparative Monte Carlo simulation study see \citet{ebner2020tests}.

In a similar vein, \citet{Betsch2019} and \citet{Betsch2020discrete} provide new characterizations of continuous and discrete parametric families of distributions through the density approach for novel tests for univariate normality \citep{betsch2020testing}, the gamma family \citep{Betsch2019a} and the inverse Gaussian law \citep{Allison2019}.  Note that other test statistics of type (\ref{eq:Tn}) based on Stein operators are implicitly proposed in tests for parametric families, although originally motivated by characterizing (partial) differential equations for integral transforms; see, for instance, \citet{Baringhaus1991} for a test of exponentiality, \cite{BH:1992} for a test of Poissonity, and \citet{HME:2012} for a test of the gamma law.

The expression in \eqref{eq:Tn} can be thought of as a weighted $L^2$-difference between the expectation of $\opsub{\widehat{\vartheta}_n}g_t$ under $P_{\widehat{\vartheta}_n}$ and $Q_n=n^{-1} \sum_{i=1}^n \delta_{x_i}$. This is in contrast with the IPMs, such as the KSD of the previous section, which measure worst-case types of differences (recall Equation \eqref{eq:IPM} which considers the supremum instead of an average). As a result, although the tests in Sections \ref{kernelSteinEmbedding} and \ref{sec:stein-oper-goodn} are both based on Stein operators, they use these in rather different manners. The tests in Section \ref{kernelSteinEmbedding} use an RKHS setting which allows for a rich set of alternative distributions.  For the tests in Section \ref{sec:stein-oper-goodn}, the benefit of considering the structure of a $L^2$-Hilbert space lies in the fact that the central limit theorem for Hilbert-space valued random elements can be exploited to derive limit distributions under $H_0$, as well as fixed and contiguous alternatives. 


\subsection{Maximum likelihood estimation and chisquare  tests}\label{sec:MLE}
With Stein's method it is possible to give explicit bounds at finite sample size $n$ to the asymptotic approximation of estimators and test statistics. The arguably most famous example is the asymptotic normal distribution for maximum likelihood estimators (MLEs) under fairly general conditions. 
    For example, in the simple case of $X_1, X_2, \ldots, X_n$ being i.i.d.\ random variables from a single-parameter distribution, then  for $Z \sim {\rm N}(0,1)$, and under classical regularity conditions,
\begin{talign}
\label{normal_approximation_univariate}
\sqrt{n\:i(\theta_0)}(\hat{\theta}_n(\bolds{X}) - \theta_0) \rightarrow_d Z, \quad \text{as $n\rightarrow\infty$,}
\end{talign}
where $\rightarrow_d$ denotes convergence in distribution. Starting with the single-parameter case, under some natural regularity assumptions which we do not detail here, \cite{anastasiou_reinert2017} obtain general bounds w.r.t.\ the bounded Wasserstein distance as follows. Let $W_n:=\sqrt{n\:i(\theta_0)}(\hat{\theta}_n(\bolds{X}) - \theta_0)$. Then the interest is to find upper bounds on $|\mathbb{E}[h(W_n)] - \mathbb{E}[h(Z)]|$, where $h\in\mathcal{H}_{\mathrm{bW}}$ as in {Remark \ref{idmlist}}. The general idea is to represent the standardized MLE {in such a way} that it contains a quantity which is a sum of independent random variables plus a term that can be controlled. The part involving the sum is handled via 
a classical {use} of Stein's Method. While the underlying random sample  $X_1, \ldots, X_n$  are assumed i.i.d.\,in \cite{anastasiou_reinert2017}, they are  locally dependent in  \cite{anastasiou2017bounds_m}. 
    
    As an illustration, consider the exponential distribution in its canonical form. The probability density function is $f(x|\theta) = \theta{\rm exp}\{-\theta x\}$ for $x > 0$  and the unique MLE for $\theta$ is  $\hat{\theta}_n(\bolds{X}) = 1/\overline{X}$, the inverse of the sample average. Then, \cite{anastasiou_reinert2017} 
    established that 
\begin{talign*}
d_{\mathrm{bW}}(\mathcal{L}(W_n), \mathcal{L}(Z))
 \leq  \frac{ 4.41456}{\sqrt{n}} + \frac{8(n+2)(1+\sqrt{n} )}{(n-1)(n-2)},
\end{talign*}
with $Z \sim {\rm N}(0,1)$ and $W_n:=\sqrt{n\:i(\theta_0)}(\hat{\theta}_n(\bolds{X}) - \theta_0)$; here, $i(\theta_0)$ is the expected Fisher information for one variable. This bound is explicit and of the order $n^{-1/2}$.
 Using the delta method  combined with Stein's method,  \cite{anastasiou_ley2017} give an explicit bound for MLEs which are a smooth function of a sum of independent terms.  This result is generalised to the multivariate case in \cite{ag20}.

Since the MLE can be used as a basis for likelihood ratio tests, which under regularity assumptions follow approximately a chi-square distribution, it 
is natural to measure the finite-sample approximation error of such tests. An explicit general bound of order $O(n^{-1/2})$ is obtained in \cite{anastasiou_reinert2018bounds} using Stein's method. 

Explicit bounds on chisquare approximations for Pearson’s chi-square test for goodness-of-fit of categorical data are obtained in \cite{gaunt2017chi}, and more generally the power divergence family of statistics in \cite{gaunt21}. 
\cite{gauntreinert21} provided explicit  bounds of the order $r/n$ to quantify the chi-square approximation with $r-1$ degrees of freedom to Friedman's statistic.

\subsection{The effect of prior choice on the posterior in Bayesian statistics}\label{sec:Bayes}
In Bayesian statistics, \cite{DF} proved that, under  certain regularity conditions and for large sample sizes,  the choice of a prior distribution gets irrelevant {for} posterior inference. With the help of Stein’s method, \cite{LRS17,GL19} complemented this result by estimating prior sensitivity  for fixed (and often small) sample sizes by quantifying the Wasserstein distance between posterior distributions arising from two distinct priors in the one-dimensional one-parameter setting. The argument was 
extended to the multivariate setting in \cite{mijoule2021stein}. 

Let us start by fixing the notation. Suppose that the observations ${X_1,\ldots, X_n}$ are i.i.d.\ from a parametric model with scalar parameter of interest which we model as some random variable $\Theta$. Now, assume we have two distinct  (possibly improper) prior densities $p_1(\theta)$ and $p_2(\theta)$ for the random quantity $\Theta$. The resulting posterior densities for $\Theta$ can be expressed as 
\begin{talign}\label{ex:ley}
p_i (\theta;x) = \kappa_i (x) p_i(\theta) \ell(\theta;x),   \  \   \  i=1,2,
\end{talign}
where $\kappa_1$ and $\kappa_2$ are 
normalizing constants. Denote by $(\Theta_1, P_1)$ and $(\Theta_2, P_2)$ pairs of random variables and cumulative distribution  functions which correspond to the densities $p_1(\theta; x)$ and $p_2(\theta; x)$, respectively. We assume that the densities $p_1(\theta; x)$ and $p_2(\theta; x)$ are \textit{nested}, {so} that the support of one is included in the support of the other. We suppose ${I}_2 \subseteq {I}_1$ which allows us to write $p_2 (\theta;x) = \frac{\kappa_2 (x)}{\kappa_1 (x)} \rho(\theta) p_1(\theta; x)$ where $\rho(\theta) = p_2(\theta)/p_1(\theta)$ is the ratio of prior densities. The key idea relies on the elementary identity
\begin{talign*}
 \frac{\frac{\mathrm{d}}{\mathrm{d}\theta}(p_2(\theta;x)f(\theta))}{p_2(\theta;x)}=\frac{\frac{\mathrm{d}}{\mathrm{d}\theta}(p_1(\theta;x)f(\theta))}{p_1(\theta;x)}+\big(\frac{\mathrm{d}}{\mathrm{d}\theta}\log(\rho(\theta))\big)f(\theta),
\end{talign*}
which is an immediate consequence of \eqref{ex:ley} and the nestedness of the densities. {This identity no longer involves the normalizing constants and it}  relates the density operators of $p_1(\cdot; x)$ and $p_2(\cdot; x)$ in such a way  that, with $f_h$ a solution  to the Stein equation $h(x) - \mathbb{E}_{X_1\sim P_1} h(X_1) = \mathcal T_1 f_h   (x)$, we get 
\begin{talign*}
\mathbb{E}_{\Theta_2 \sim P_2} [h(\Theta_2)] - \mathbb{E}_{\Theta_1 \sim P_1} [h(\Theta_1)] & =\mathbb{E}_{\Theta_2 \sim P_2}  \left[\mathcal{T}_1 f_h(\Theta_2)\right]   =\mathbb{E}_{\Theta_2 \sim P_2}  \left[(\mathcal{T}_1 - \mathcal{T}_2) f_h(\Theta_2)\right] \\
& = \mathbb{E}_{\Theta_2 \sim P_2}  \left[\frac{\mathrm{d}}{\mathrm{d}\theta}\log(\rho(\theta))\big|_{\theta = \Theta_2} f_h(\Theta_2 )\right] \end{talign*}
(the second equality holds because, by definition,  $\mathbb{E}_{\Theta_2 \sim P_2}  \left[  \mathcal{T}_2 f_h(\Theta_2)\right] = 0$). Thus, bounding an IPM generated by some class $\mathcal H$ between $\Theta_2$ and $\Theta_1$ can be achieved by bounding  $\mathbb{E}_{\Theta_2 \sim P_2}  \big[\frac{\mathrm{d}}{\mathrm{d}\theta}\log(\rho(\theta))\big|_{\theta = \Theta_2} f_h    (\Theta_2 )\big]$ over all   $h \in \mathcal H$.

For the sake of illustration, consider normal   data with fixed variance $\sigma^2$, and  the mean   {the}  parameter of interest. \cite{LRS17}  compare a normal ${\rm N}(\mu,\delta^2)$ prior for the location parameter (the conjugate prior in this situation) with a  uniform prior.
They  bounded the  Wasserstein distance between the resulting posteriors $P_1$ and $P_2$ by
\begin{talign*}
\frac{\sigma^2}{n\delta^2+ \sigma^2}\big| \overline{x}-\mu \big| \leq d_{\mathrm W} (P_1, P_2) \leq \frac{\sigma^2}{n\delta^2+ \sigma^2}\big| \overline{x}-\mu \big| + \frac{\sqrt{2}}{\sqrt{\pi}} \frac{\sigma^3}{n\delta \sqrt{n \delta^2 + \sigma^2}}
\end{talign*}
with $\overline{x} = n^{-1} \sum_{i=1}^n x_i$ the sample average. Both bounds are of the order of $O(n^{-1})$ and are easily interpreted: the better the initial guess of the prior, meaning here of the location, the smaller the bounds and  hence the smaller  the influence of the prior.

\section{Conclusion}\label{sec:conclu}

The goal of this paper is to highlight some recent  developments in computational statistics that have been accomplished via tools inherited from Stein's method.  
Moreover, this paper illustrates that there is considerable scope for more interplay between the research strand on how to set up Stein operators and that of devising computable Stein discrepancies and related algorithms. For example, for a given target distribution, it is  mostly an open problem which Stein operator and class to choose so as to obtain a computable Stein discrepancy which is most useful for the problem at hand. This answer may 
differ depending on whether we want to construct a hypothesis test, develop a sampling method, or measure sample quality;
a step in this direction is taken in \cite{xu2022standardisation}. Section \ref{sec:goodness-of-fit} highlights how Stein's method can be brought to fruition not only to devise estimators but also to quantify their behaviour. There is plenty of scope for analysing the procedures and estimators from Sections  \ref{sec:estimators} and \ref{kernelSteinEmbedding} to obtain quantitative bounds on their performance.

The list of results given in this paper are but a mere sample of the ongoing activity  in this newly established   area of research at the boundary between probability, functional analysis, data science and computational statistics. 
For instance, Stein's method has been used  for designing sampling-based algorithms for non-convex optimization \citep{ErdogduMaSh2018}, or for learning semi-parametric multi-index models in high dimensions \citep{yangetal2017}. In Bayesian statistics, Stein discrepancies have been used as variational objectives for posterior approximation \citep[e.g.,][]{ranganath2016operator,hu2018stein,fisher2020measure}. 

 A complete exhaustive description  of all recent developments in this area is an impossible task within the constrained space of a review paper such as this one. Yet, we hope that the range of problems which are addressed in this paper show the versatility of Stein's method, and the promise that it holds for further exciting developments.

\hfill

\noindent \textbf{Acknowledgments}:

AA was supported by a start-up grant from the University of Cyprus. AB was supported by  the UK Defence Science and Technology Laboratory (Dstl) and Engineering and Physical Research Council (EPSRC) under the grant EP/R018413/2. FXB and CJO were supported by the Lloyds Register Foundation Programme on Data-Centric Engineering and The Alan Turing Institute under the EPSRC grant EP/N510129/1.  AG was supported by the Gatsby Charitable Foundation. RG was supported by a Dame Kathleen Ollerenshaw Research Fellowship. FG and CL were supported by a BOF Starting Grant of Ghent University. QL was supported in part by NSF CAREER No. 1846421.  GR was supported in part by EP/T018445/1 and EP/R018472/1. YS was supported in part by CDR/OL J.0197.20 from FRS-FNRS. The authors thank the Editor, Associate Editor and two anonymous reviewers for helpful comments that led to a clear improvement of the presentation of this paper.

\footnotesize

\bibliographystyle{apalike}
\bibliography{bibliography-abbrev}

\end{document}